\documentclass[onecolumn]{aastex63}
\shorttitle{Revised Cepheid Calibration}
\shortauthors{Madore, Freedman, et al.  }
\graphicspath{{./}{figures/}}
\begin{document}
\title{\bf Revising the Milky Way Cepheid Calibration: \\Quantifying and Correcting for Previously Undetected Distance Modulus Errors in the Gaia-based Multi-Wavelength Period-Luminosity Relations.}
\begin{abstract} 
We examine the multi-wavelength period–luminosity–color (PLC) relations for Cepheid variables in the Large and Small Magellanic Clouds and the Milky Way. From first-principles stellar physics, the luminosity of a Cepheid is determined by its radius and surface temperature, yielding a fundamental PLC relation whose observational proxies are pulsation period and intrinsic color. Using Cepheids in the Magellanic Clouds, we show that the PLC relation recovers the known geometries and line-of-sight tilts of their disks, confirming its ability to detect true distance-modulus variations that are achromatic and consistent across all filters. Surprisingly, for Milky Way Cepheids with individually determined reddenings and HST and Gaia parallaxes, the residuals from multi-wavelength PL fits are also found to be achromatic, identical in sign and amplitude across all passbands, in this case indicating that parallax errors are the dominant source of scatter. Applying bandpass-averaged corrections to individual Cepheids recovers the theoretically expected wavelength-dependent narrowing of the instability strip, and results in revised parallaxes with a median improvement in precision of roughly a factor of two. In addition, they show no statistically significant correlation with metallicity over the range –0.2 $<$ [Fe/H] $0<$ +0.5 dex. The final extinction- and reddening-corrected PLC relation yields an rms scatter of ±0.04 mag, corresponding to $\sim$2\% precision in distance per star. Use of a physically grounded PLC will provide a more robust foundation for the Cepheid-based extragalactic distance scale and the determination of the Hubble constant.

\end{abstract}

\author[0000-0003-3431-9135]{\bf Barry F. Madore} 
\affil{The Observatories, Carnegie
Institution for Science, 813 Santa Barbara St., 
Pasadena, CA ~~91101, USA}
\affil{Department of Astronomy \& Astrophysics, University of Chicago, 5640 South Ellis Avenue, Chicago, IL 60637, USA}
\email{barry.f.madore@gmail.com} 

\author[0000-0003-3431-9135]{\bf Wendy~L.~Freedman}
\affil{Department of Astronomy \& Astrophysics, University of Chicago, 5640 South Ellis Avenue, Chicago, IL 60637, USA}
\affil{Kavli Institute for Cosmological Physics, University of Chicago,  5640 S. Ellis Ave., Chicago, IL 60637, USA}
\email{wfreedman@uchicago.edu}


\section{Background Introduction}
The local extragalactic distance scale has long relied upon one method, the Cepheid PL relation, as its base. 
As such, the Cepheid distance scale has been at the center of all of the ``crises'' confronting the size of the universe since the time of Curtis, Shapley, and shortly thereafter Hubble himself (Smith, 1982).
The Cepheid distance scale has gone through numerous  revisions each at the factor-of-two level, starting with Baade (1956) correcting Hubble, followed by Sandage \& Tammann (1982) and de Vaucouleurs (1983) debating each other over values of $H_o = $ 50 or 100 km/s/Mpc, respectively. 
The latter, extended debate went unresolved until the {\it Hubble Space Telescope} (HST) was launched, allowing the determination of a Cepheid-based distance scale tying in to distant supernovae and four additional secondary methods, giving $H_o = $ 72 km/s/Mpc with a 10\% uncertainty (Freedman et al. 2001). However, even that ``resolution'' was short-lived. 
The {\it Planck} mission, and its cosmological modeling of the power spectrum of the cosmic microwave background (CMB) radiation did not confirm the Cepheid distance scale. This led to the most recent ``crisis in cosmology'', now commonly referred to as the ``Hubble Tension'' between the standard Lambda Cold Dark Matter (Lambda-CDM) model  and the Cepheids. 
More than a thousand publications, attempting to modify/augment Lambda-CDM have so far failed to find a theoretical solution moving the cosmological theory towards better agreement with the Cepheid distance scale (di Valentino et al. 2021).

 In the following we explore both the current limits and future promise in precision of the Cepheids, exploiting multi-wavelength data needed later to address the systematic effects of line-of-sight extinction/reddening to individual stars. In addition, we investigate  residual metallicity effects.  
\section{Back to Basics}
 First-principle physics predicts that all stars (not just Cepheids) are described by two physically independent parameters: a size, $R$ and a surface temperature, $T$. 
 Simple geometry gives the total radiating surface area, A for a spherical star: $A  = 4 \pi R^2$.
 And for blackbody radiation having a temperature of $T_{eff}$, the Stefan-Boltzmann equation gives a surface brightness of $\sigma T_{eff}^4$, where the proportionality constant,  $\sigma$
 is the Stefan-Boltzmann constant.
 The total (bolometric) luminosity of the star is then the product of these two terms, yielding $L = 4 \pi R^2 \times\sigma T_{eff}^4$ ~... (1), or expressed in conventional astronomical units of magnitude, $M_{bol} = -5~ log R - 10~ log T_{eff} + C$~~~... (2). 
 Their observational counterpart is the Period-Luminosity-Color (PLC) relation for Classical Cepheids, a wavelength-dependent example of which is $M_{V} = \alpha~log P + \beta~(B-V)  + \gamma $~~... (3)\footnote{See Sandage \& Tammann 1968, Sandage 1972 for early and influential examples of this line of argument, from the observational side, and then {\it a fortiori} by Bono \& Marconi 1999, and Bono, Caputo, Castellani, et al. 1999 from the theoretical modeling side, including an early discussion comparing the relative advantages of the Wesenheit and PLC formulations.} 

 However, it should be pointed out that nowhere in Equation 1 is there any explicit mention of the period of oscillation of the star, which may or may not be manifest observationally. 
Likewise in Equation 3 there is no mention of the radius of the Cepheid, which for any star is a very difficult property to measure in all but the rarest of circumstances, for instance, in edge-on eclipsing stellar systems. 
Nevertheless, an apparently simple (and ultimately misleading) invocation of the 
$P\sqrt{\rho} = Q ~~... ~(4)$ relation for oscillating bodies was used, not only by Sandage half a century ago, but also in one of this century's most recent textbooks on stellar structure (Lamers \& Levesque 2017). There, the authors swap in period for radius using the following re-grouped version of the $P\sqrt{\rho}$
relation (Equation 4) giving $R = P^{2/3}(M/4 \pi Q)/3)^2)^{1/3}$. However, this simplification does not allow for the fact that $M$ is not a constant for Cepheids in the instability strip. Furthermore, it is sometimes stated that the Cepheid instability strip is nearly vertical in the color–magnitude diagram, which would in principle permit the elimination of the color term in the Period–Luminosity–Color (PLC) relation. In practice, however, Cepheids are known to obey a well-defined Period–Color relation, and the assumption of a vertical instability strip is not supported observationally. Neglecting the color dependence therefore removes a physically meaningful parameter and obscures the role of temperature variations, which are fundamentally constrained by stellar structure and radiative physics.
These assumptions lead to a simplified period–luminosity (PL) relation and can give the impression of a physically complete one-parameter description, commonly referred to as the Leavitt Law. We instead argue that retaining the full PLC formulation provides a more physically motivated and internally consistent framework, and we demonstrate below that it offers the most powerful route to a high-precision, multi-wavelength Cepheid distance scale.
In the process, we will show that a full appreciation of the PLC allows us to discover and correct for parallax errors in the most recent {\it Gaia} and {\it HST} calibrations of the Milky Way Cepheid distance scale, further increasing the demonstrable precision of the entire enterprise.

\section{STEP 1 : The Large Magellanic Cloud}
We start the process by analyzing the multi-wavelength data for Cepheids in the Large Magellanic Cloud, where it is already established that line-of-sight, back-to-front differences in distances to {\it individual Cepheids} can be measured across the face of the galaxy (e.g., Caldwell \& Coulson 1986, Haschke et al. 2012).
We show that the PLC can be utilized to detect and measure this physical scatter in the true distance moduli to individual Cepheids in the LMC, and that they confirm previous studies of the orientation and line-of-sight tilt of the disk of the LMC.
Unlike other known magnitude residuals in the multi-wavelength PL relations that are wavelength-dependent (such as those arising from reddening or intrinsic color, driven by astrophysics, and that scale predictably with wavelength),  distance-modulus errors are achromatic and identical across all filtered PL variants.
Thus, for any Cepheid that is physically displaced by $\Delta$ from the average distance modulus $<\mu>$, that difference in magnitude space will show up with the same amplitude and the same sign, in whatever wavelength PL relation it is observed in.
At the other extreme, photometric measurement errors are, by definition, random with a zero mean, and do not correlate in size or sign from one wavelength to the next.

\begin{figure*} 
\includegraphics[width=18.0cm, angle=-0]{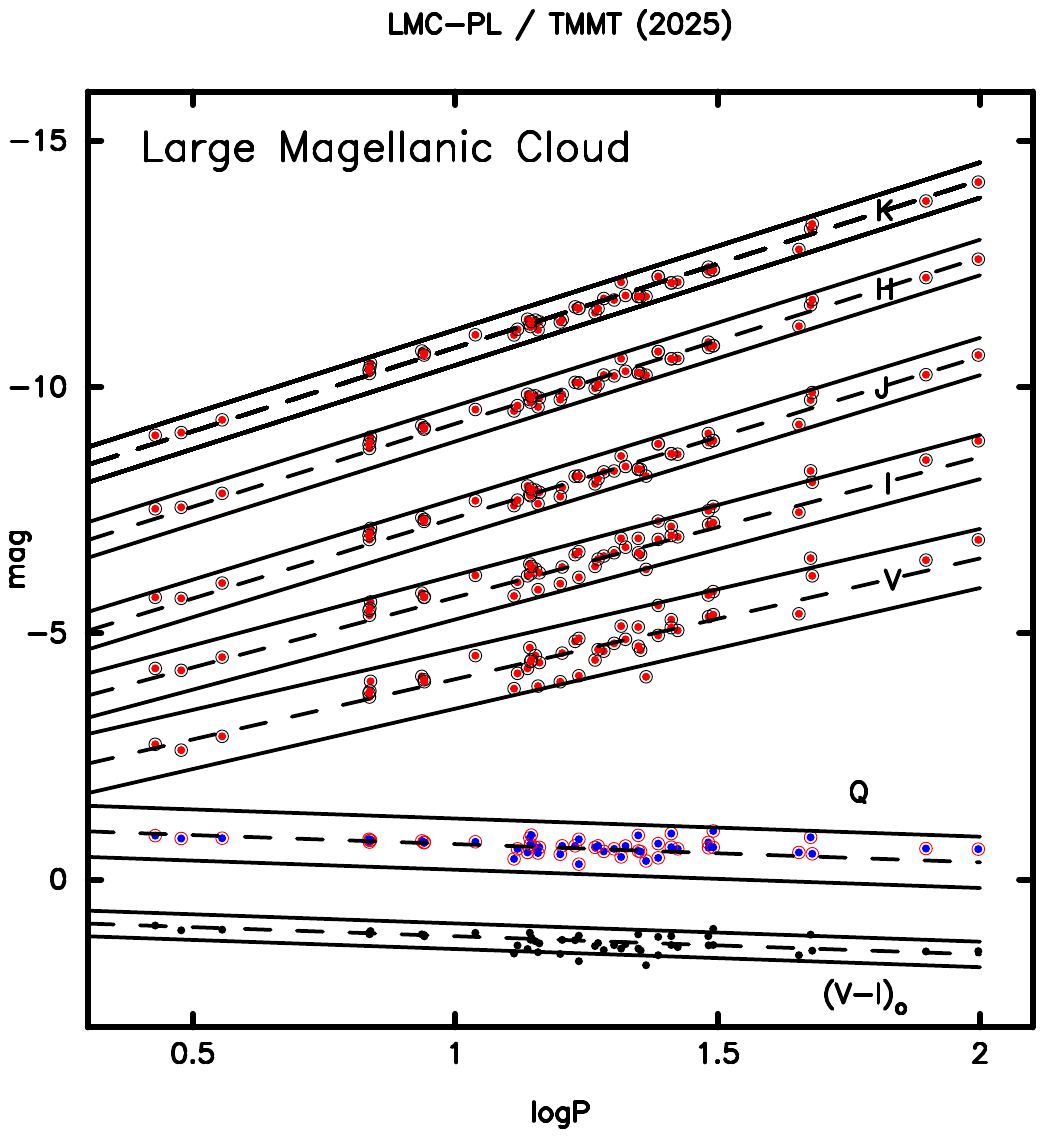} 
\caption{Multi-wavelength period-luminosity relations (upper, red points) and period-color relations (lower, purple and black points) for LMC Cepheids.
}
\end{figure*}

\begin{figure*} 
\includegraphics[width=18.0cm, angle=-0]{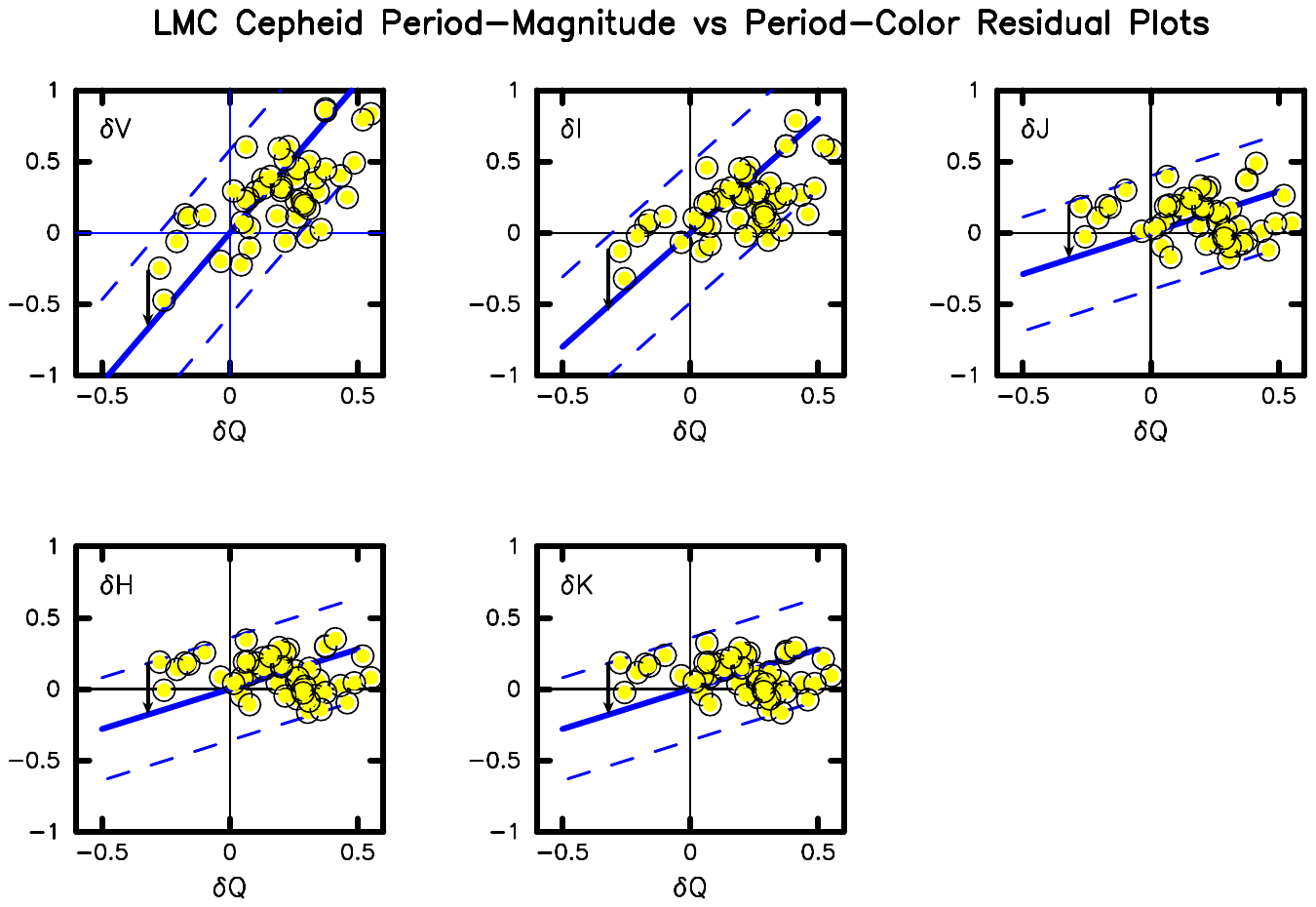 } 
\caption{VIJHK residuals from their respective PL relations plotted against the corresponding reddening-free PC Q-color residuals. The expected correlations for each bandpass are shown by solid blue lines, with $\pm$ two-sigma bounds given by parallel broken blue lines. A single downward-pointing arrow indicates one example of an achromatic offset noted for one of the many such correlated residuals easily tracked across the five panels.
}
\end{figure*}
\begin{figure*} 
\includegraphics[width=18.0cm, angle=-0]{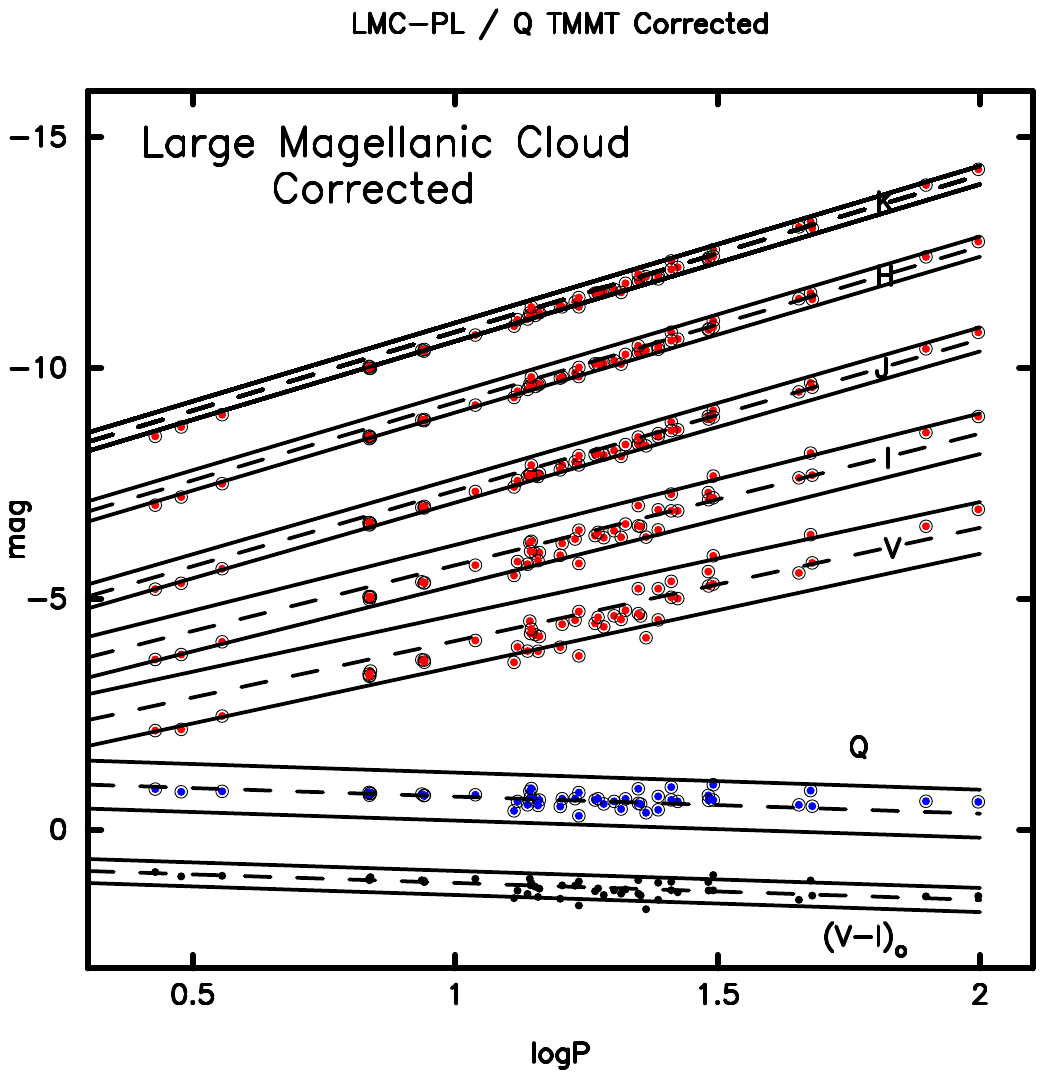} 
\caption{Same as Figure 1, expect the PL data are corrected for tilt-induced magnitude deflections. Colors are unaffected by distance errors and/or corrections.
}
\end{figure*}

\begin{figure*} 
\includegraphics[width=18.0cm, angle=-0]{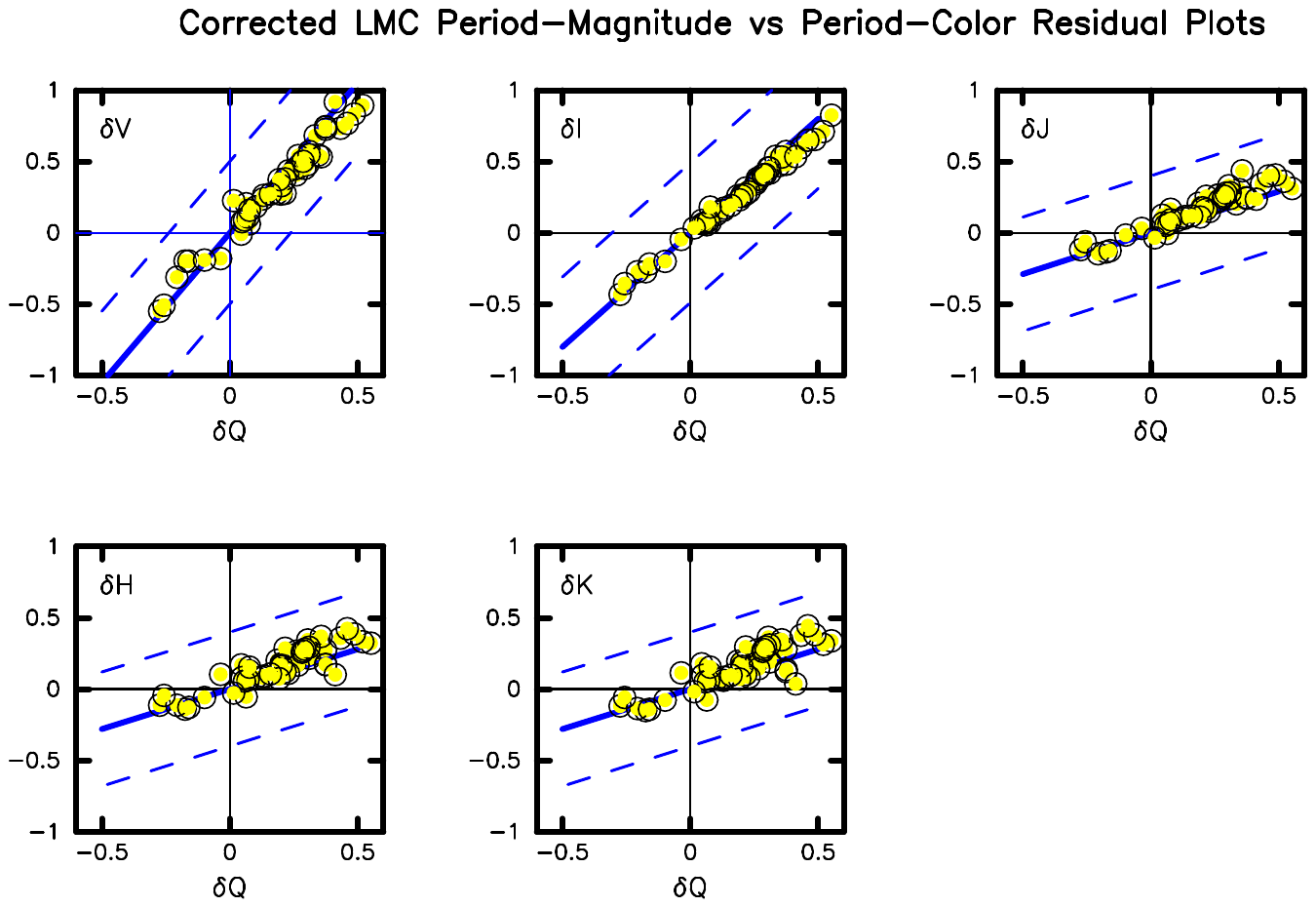}
\caption{The same as Figure 2 except the data plotted here have been corrected for the individual, achromatic (delta distance modulus) offsets by the application of one number to each Cepheid at each of the wavelengths.
The data are now extremely well fit to the expected (blue) trends of magnitude with intrinsic color.
}
\end{figure*}

Figure 1 shows the multi-wavelength PL relations for sample of Cepheids in the LMC, as published by Breuval et al. (2021). Here, we down-selected from the total known population of LMC Cepheids so as to be in the same general area as the detached eclipsing binary (DEB) stars used by Pietrzyński, et al. (2019) to determine a geometric distance to that central region of the LMC. As can be easily seen, there is a modest but monotonic decrease in the widths of the PL relations going from the optical (VI) to the near-infrared (JHK). Furthermore, there is a strong correlation from one PL relation to another visible in the relative positions of individual Cepheids across the various bandpasses.

At the bottom of the plot are two colors, $(V-I)_o$ in black and, shown in blue, a reddening-free, Qualine color, (Hiltner \& Johnson 1956) defined as $Q = (V-I) - X (I-K)$, where $X = E(V-I)/E(I-K)$. While there is good correspondence between the positions of individual Cepheids with respect to the dashed trend lines; the same cannot be said for the detailed comparison of color scatter and magnitude scatter, which Figure 2 explores in more detail.

If the expectations of a PLC relation are to be realized in fact, it is hoped that color differences across the Cepheid instability strip will be directly driving magnitude differences in each of the corresponding PL relations. All five of the sub-panels in Figure 2 are rather disappointing in that regard, given the high quality of the photometry upon which they are based, and which is evident in the ($V-I$) colors plotted at the bottom of Figure 1. The expected correlations are shown in Figure 2 as solid blue lines in each of the sub-plots. This figure leads to one of the major findings in this paper. As illustrated for one of the points to the far left of each of the five sub-plots there is an arrow pointing down from the Cepheid to the fiducial line. Regardless of the slope of the line, which decreases systematically from  V to K, the length of that arrow is virtually the same across all of the plots. The same concordance of magnitude residuals from plot to plot is true for all of the Cepheids.

Next we proceeded to determine the magnitude offsets on a star-by-star basis, and then averaged those (achromatic) offsets across the bands. We then subtracted the averaged offset from the original magnitudes and replot the PL relations, shown now in Figure 3. The result of applying one (wavelength independent) correction for each Cepheid, regardless of the bandpass being considered, is immediately obvious to the eye, when comparing Figures 1 and 3.

Because of the achromatic nature of the magnitude deviations revealed in the residual-residual plots, discussed above, we interpret the averaged offsets as being due to true differences in the distances of each of those Cepheids with respect to the averaged difference to the LMC, defined by the ensemble, a result consistent with the known tilted plane of the disk of the LMC  with respect to our line of sight. It is not our intent, at this time, to derive the parameters of the LMC tilt and position angle 
from this restricted dataset, but rather simply to show that our method of reducing "scatter" has identified known physical differences in distances. It also allows us to calibrate the dependence of the remaining scatter on intrinsic color. The resulting transformation of Figure 2 is shown in Figure 4. The recovery of the geometry of the LMC  using our averaged magnitude residuals, is shown in Figure 5.   

The scatter in the correlations of magnitude and color seen in the sub-panels of Figure 4 
are, respectively,  $\sigma_V = \pm 0.094, \sigma_I = \pm 0.044, \sigma_J = 0.094, \sigma_H = \pm 0.098, \sigma_K = \pm 0.107$ mag. 
Using the PLC formalism is thereby delivering distances to individual Cepheids in the LMC at the $\pm$ 5\% level per star.

For completeness, we next run the same test on Cepheid data (van der Marel \& Cioni 2001) in the Small Magellanic Cloud.
\vfill\eject
\begin{figure*} 
\includegraphics[width=18.0cm, angle=-0]{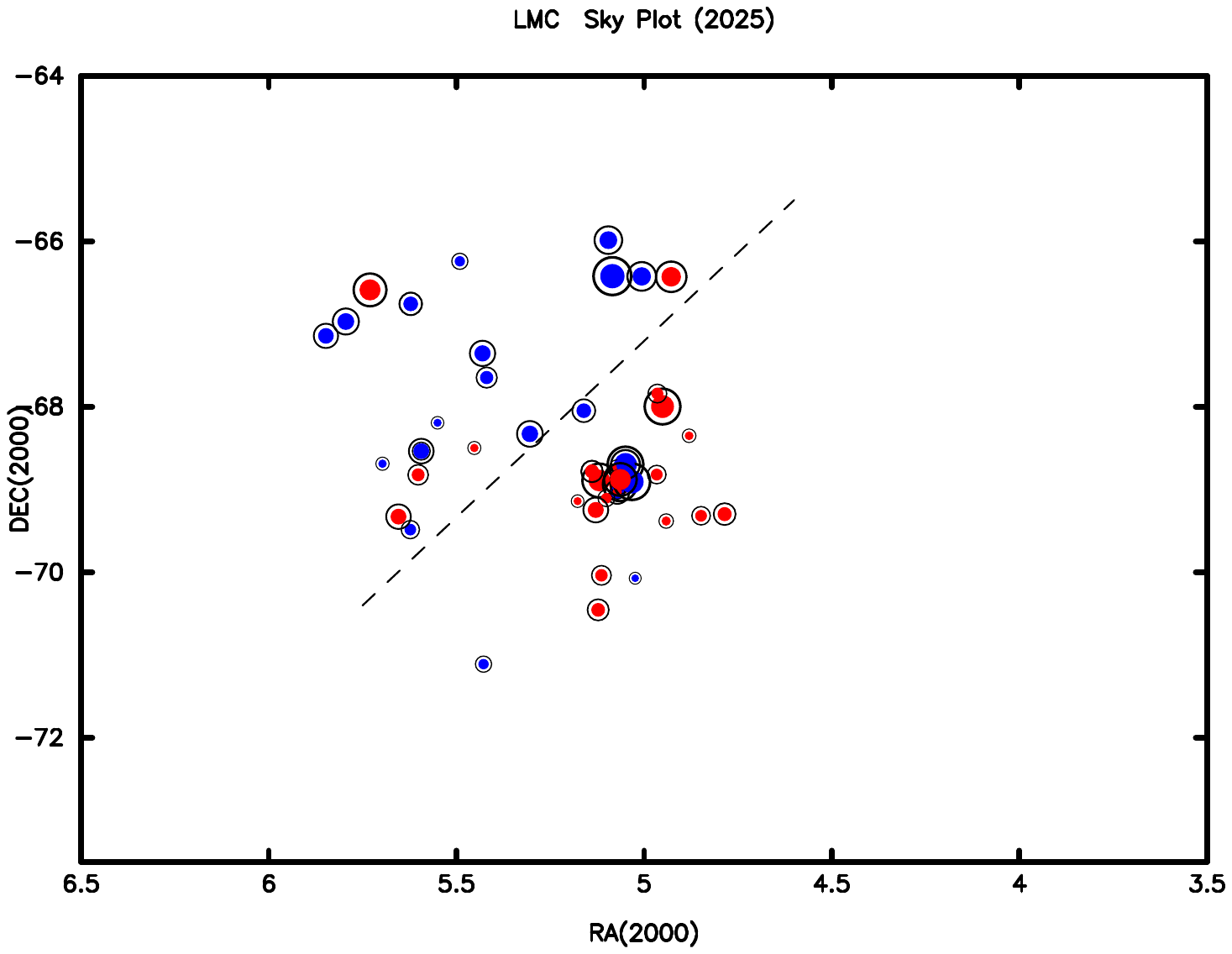}
\caption{Sky plot of our sample of Cepheids in the LMC, color-coded by their line-of-sight distances relative to the adopted distance to the galaxy as a whole. Blue points grow in size as their distances behind the LMC increase; red points are nearer. The dashed line is the approximate line of nodes for the LMC (e.g., Persson et al. 2004).}
\end{figure*}
\clearpage

\section{Step 2: The Small Magellanic Cloud}
The Small Magellanic Cloud (SMC), is somewhat (half a magnitude) more distant than the LMC and even more widely dispersed across the sky, presumably due to its interaction with the LMC, specifically, and the Milky Way more generally (e.g., Lucchini 2024).
The data used here are those published by Breuval et al. (2021), again restricting it to a small central region of the SMC, chosen to coincide with the DEB population used to establish a geometric distance to the core of the SMC (Graczyk 2020).

The PL plots (Figures 6 \& 8), the residual-residual magnitude-color plots (Figures 7 \&  9) and the on-sky geometry plot (Figure 10) parallel the discussion given in the previous section on the LMC, the main differences being the larger size of the SMC sample and the more compact central view of the parent galaxy itself in the case of the SMC.

Figure 6 shows the published optical (VI) and near-infrared (JK) PL relations for the SMC based on the above data. Also shown at the bottom of the plot are (1) the mean foreground reddening-corrected $(V-I)_o$ Period-Color relation (blue points) and (2) the Qualine, Q = (V-I)-R$\times$(I-K), Period-Intrinsic Color relation (black points).
The very close star-by-star correspondence of the individual data points in these two PC diagrams is a clear indication of the low amount of {\it differential} reddening along this particular line-of-sight to the SMC, and is a testimony to the high precision of the magnitudes and colors of these Cepheids under study. The four PL relations show a slow but noticeable decrease in their widths as a function of increasing wavelength, as expected.

Despite the high precision of the SMC Cepheid photometry noted above, our expectations of a close correspondence of the intrinsic color residuals, $\delta Q$ with the magnitude residuals, $\delta V$, $\delta I$, $\delta J$, $\delta K$ and $\delta W$ are not met, and at first glance, appear to be random-noise, scatter plots. Closer examination suggests otherwise. 

The same extraction of achromatic scatter in the vertical, $\delta$ ~mag direction across these residual plots paints the same picture as revealed in the LMC study in the section above. Back-to-front differences in the true distances for individual stars are accounting for the bulk of the scatter seen in the sub-panels of Figure 7. Applying these distance modulus corrections to each star individually in each of the PL relations results in Figure 8. These four PL relations are significantly impacted by the single-number corrections uniquely applied to each star across all bands.
Specifically, the scatter in each of the PL relations has decreased, with the percentage decrease being a strong function of increasing wavelength. The resulting PL relations have one-sigma scatters of $\pm$0.19, 0.14, 0.09 \& 0.08~mag in V,I,J \& K bands, respectively.

Perhaps even more importantly, the astrophysically predicted correlation of magnitude and color deviations from the distance modulus corrected PL relations are revealed at strikingly high precision in all bands, as shown in Figure 9.

As in the case for the LMC we close out this section with a spatial plot of the derived differential distance moduli across the face of the SMC (Figure 10). Once again, there is a trend of the more distant Cepheids (red points) to the south-west as roughly indicated by the slanting black line crossing the data. This corresponds well with what has previously been found by Smitha \& Annapurni (2012) in their much more widely based study of the 3-dimensional structure of the SMC when they state ``that the inner SMC is slightly elongated in the NW-SE direction." We take this as another external confirmation of our associating the achromatic magnitude residuals with true differences in the back-to-front distances, and a modest elongation of individual Cepheids across even this small portion of the central region of the SMC.

The scatter in the correlations of magnitude and color seen in the sub-panels of Figure 9 
are, respectively,  $\sigma_V = \pm 0.131, \sigma_I = \pm 0.052, \sigma_J = \pm 0.098,  \sigma_K = \pm 0.082$ mag. 

Using the PLC formalism is thereby delivering distances to individual Cepheids in the SMC at the $\pm$ 4-6\% level per star.

With that said, we now move forward to analyzing the same residuals found in a variety of calibrations of multi-wavelength PL relations for Milky Way Cepheids,  {\it this time associating the achromatic deviations with corrections that can be attributed to uncertainties in the parallaxes from Gaia, HST and/or the Baade-Wesselink method.}

\begin{figure*} 
\includegraphics[width=18.0cm, angle=-0]{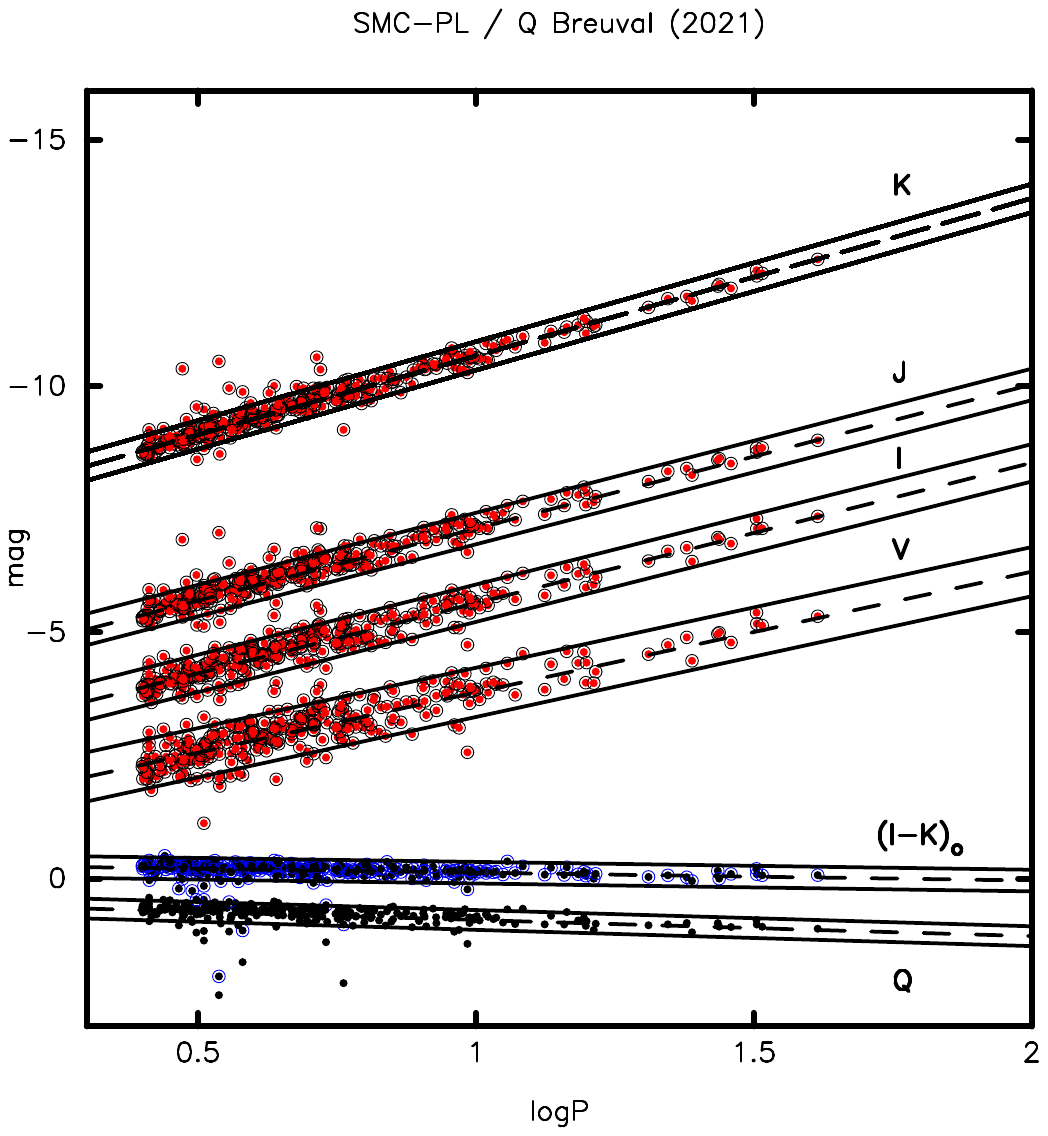} 
\caption{Multi-wavelength Period-Luminosity relations (upper, red points) and Period-Color relations (lower points) for SMC Cepheids. 
}
\end{figure*}
\begin{figure*} 
\includegraphics[width=18.0cm, angle=-0]{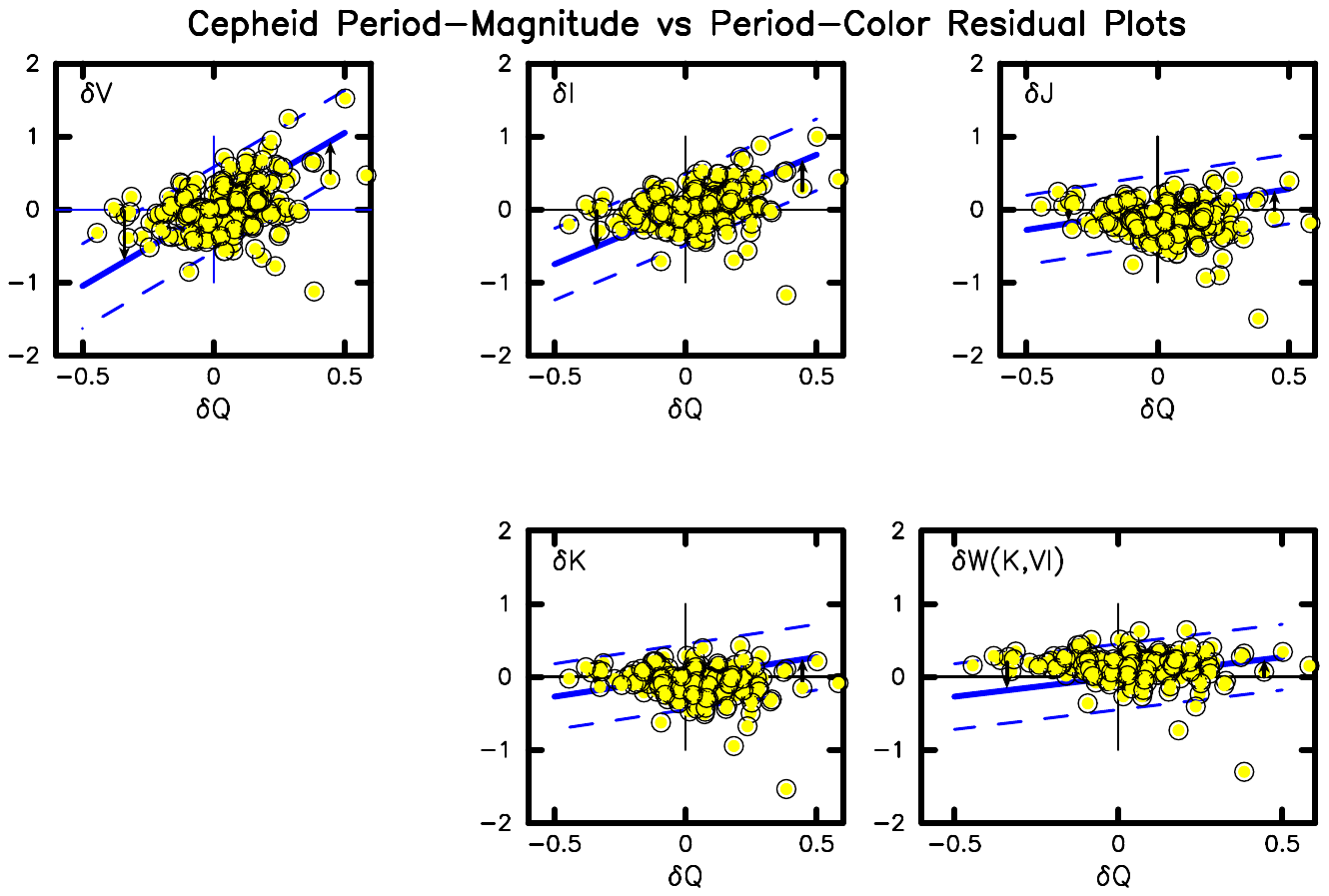} 
\caption{ Details are the same as in Figure 1. 
}
\end{figure*}

\begin{figure*} 
\includegraphics[width=18.0cm,angle=-0]{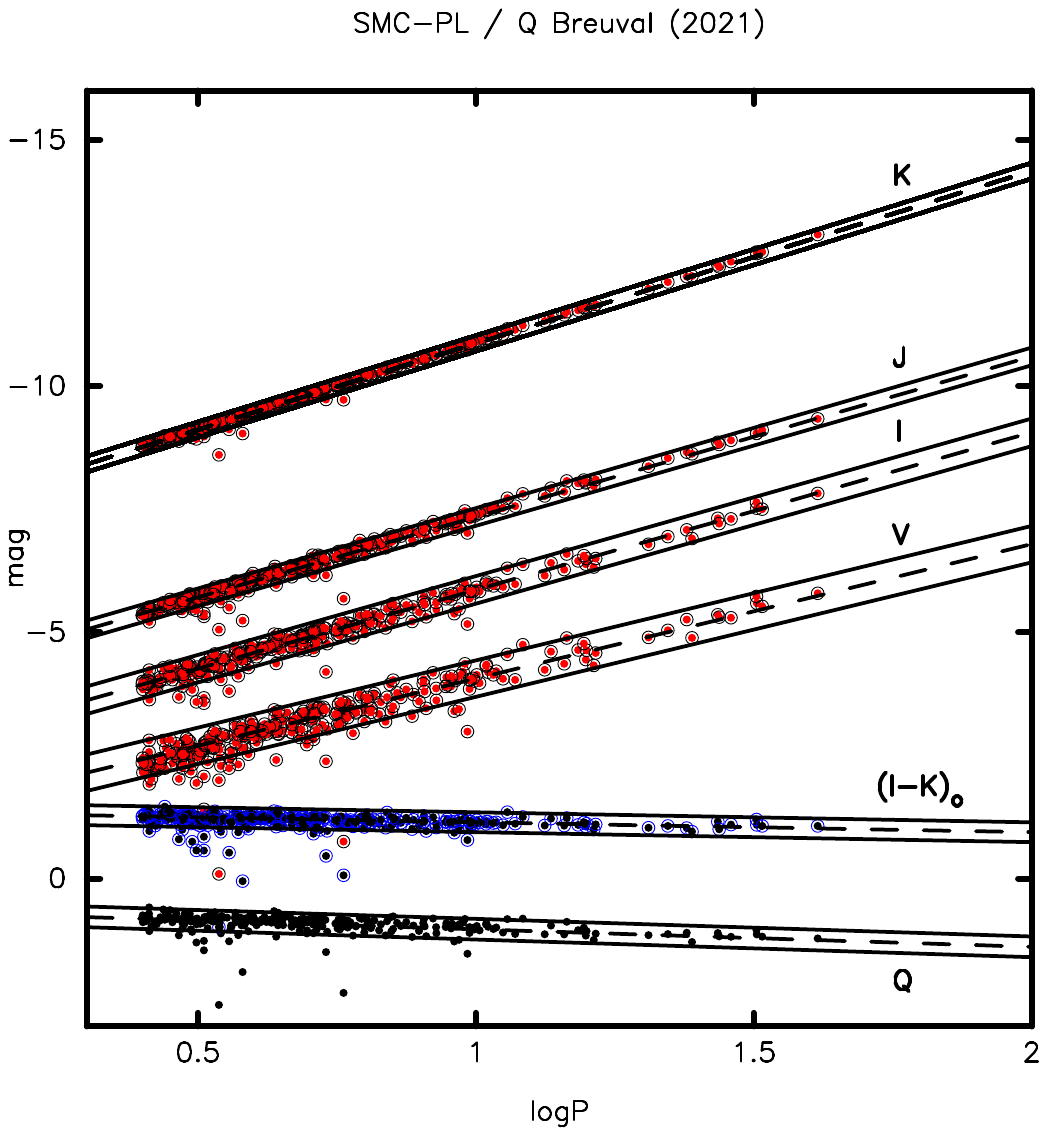} 
\caption{Multi-wavelength Period-Luminosity relations, after applying back-to-front tilt correction (upper, red points) and Period-Color relations (lower points) for SMC Cepheids. 
}
\end{figure*}

\begin{figure*} 
\includegraphics[width=18.0cm, angle=-0]{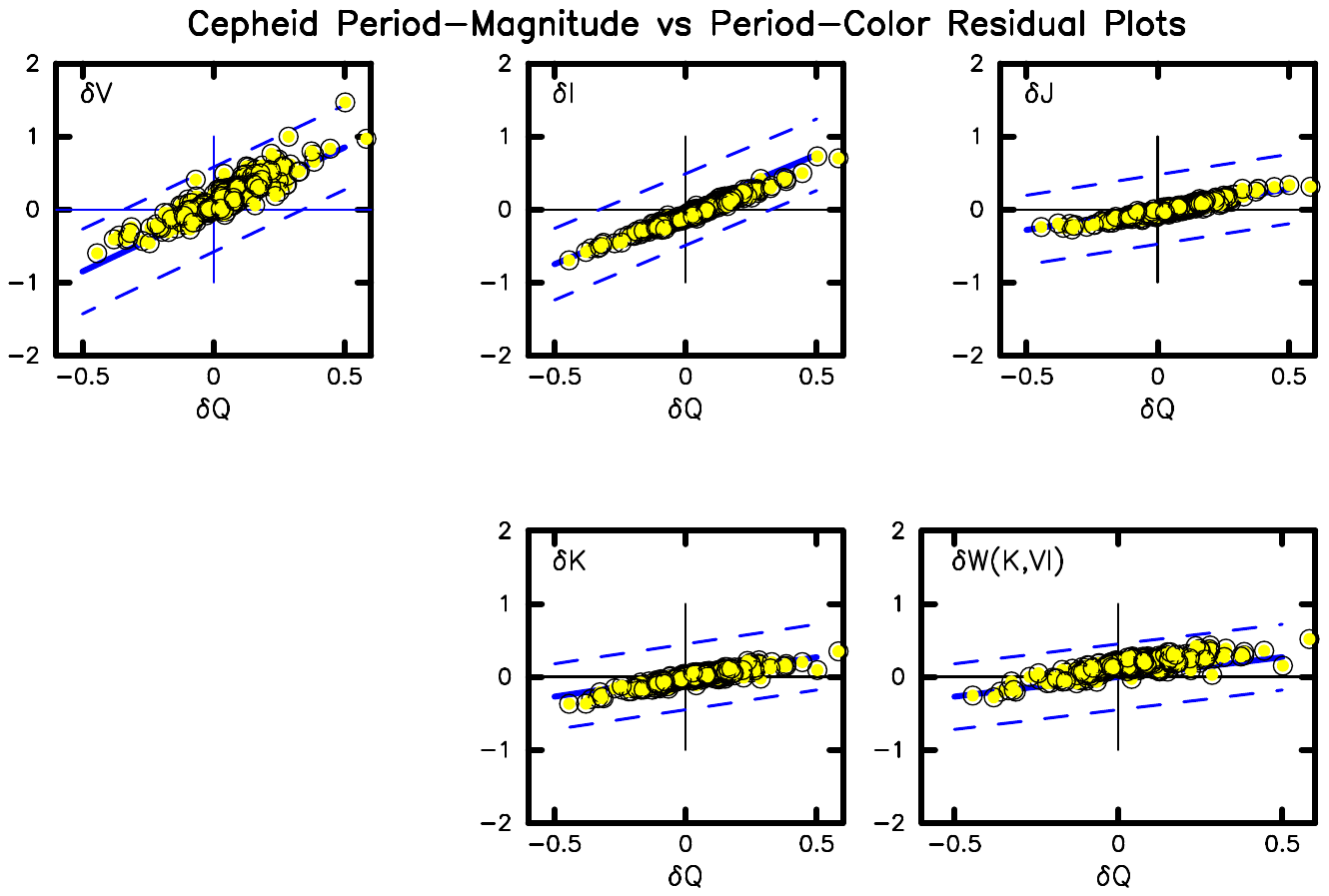} 
\caption{SMC corrected residuals. Detailsare the same as Figure 1.
}
\end{figure*}
       
\begin{figure*} 
\includegraphics[width=18.0cm, angle=-0]{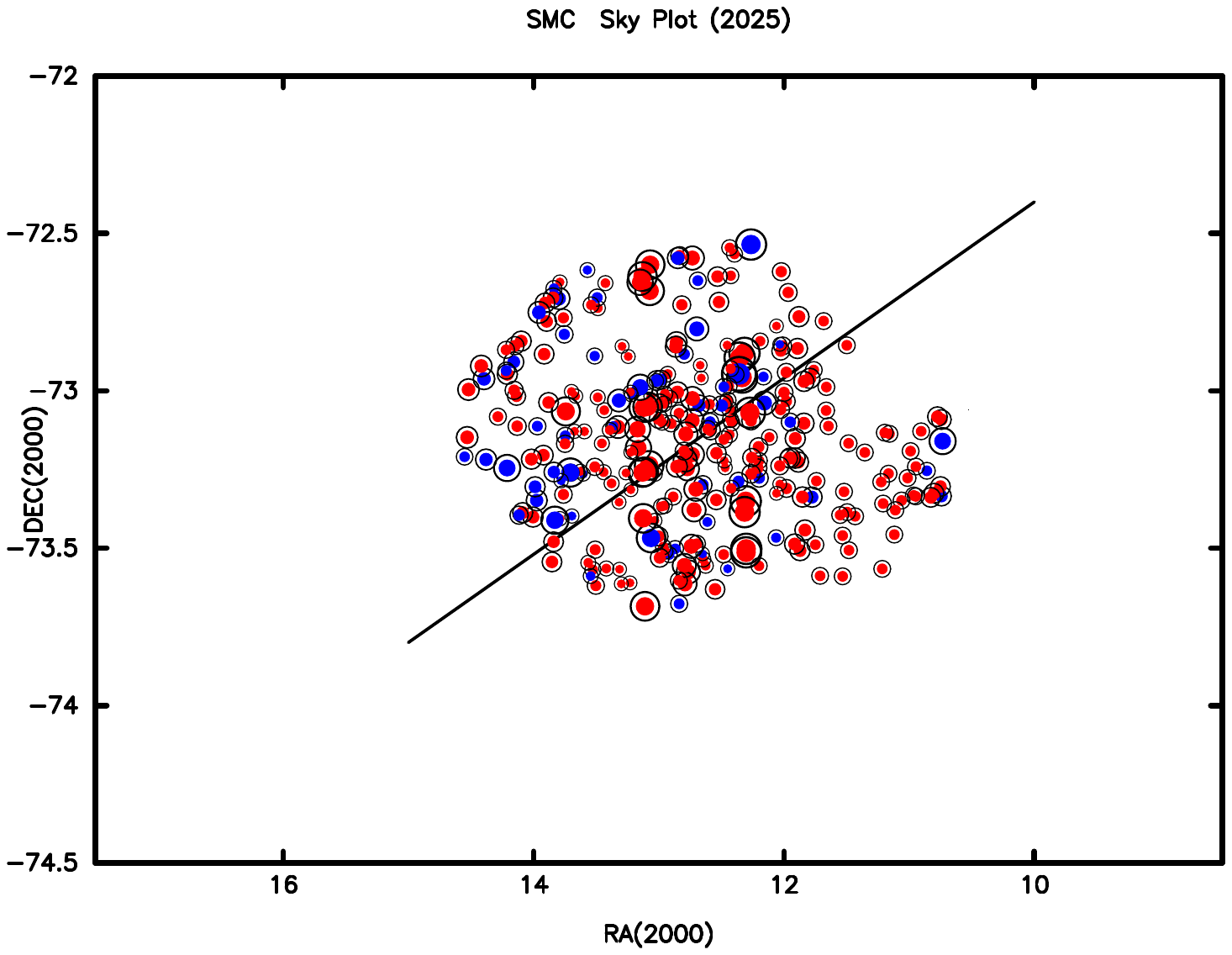} 
\caption{SMC back-to-front distance residuals.
}
\end{figure*}

\clearpage
\section{STEP 3: A Milky Way Field Sample (Breuval et al. (2021)}

Gaia parallaxes and multi-wavelength photometry for Milky Way field Cepheids have been compiled by Breuval et al. (2021), who also incorporated  high-precision metallicities from Luck (2018). For individual Cepheids, Gaia parallaxes have errors that are thought to be  driven by apparent magnitude, color, and position on the sky, while globally having a zero-point offset with respect to quasar parallaxes which, given their cosmological distances, should be (but are not observed to be) statistically zero (see Lindegren et al. 2021). Our method of analyzing the multi-wavelength PL residuals, as described above for the LMC and SMC, should be able to detect and measure differential errors in the Gaia parallaxes, seen as achromatic residuals in the magnitude-intrinsic color diagnostic plots. In the Magellanic Cloud studies we were detecting physical offsets in distance (that are wavelength independent) when compared to the average distance of each of the corresponding Magellanic Cloud populations. In the Milky Way example we will be measuring parallax errors (again, that will be manifest as individual achromatic distance modulus offsets) from the mean Gaia zero point used for the total ensemble field MW Cepheids in the Breuval sample.

There is an indication in a visual inspection of the multi-wavelength PL relations in Figure 11 (based on the absolute magnitudes published by Breuval et al. (2021) that something subtle is amiss. We note the dispersions of the long-wavelength JHK PL relations that, as compared to the Magellanic Cloud relations have dispersions a factor of two larger. Given the high precision of the photometry collected for the mean magnitudes (and colors, as seen in the bottom of the plot) of these Cepheids we should expect to see well-defined correlations of the period-luminosity residuals with respect to the residuals derived from any intrinsic period-color relation for those same stars. In that regard Figure 11 fails to meet those expectations, where increased/decreased intrinsic color (i.e. temperature) is physically responsible for increased/decreased luminosity
away from the mean line at fixed period. Something in this Milky Way sample is destroying that physically required correlation.

 A casual inter-comparison of the scatter in each of the residual-color vs residual-luminosity plots in the mosaic of Figure 12 gives an immediate answer as to the origin of the inflated scatter. Consider the luminosity deviation of a given Cepheid, above and/or below the expected luminosity relation as a function of the color deviation (shown as the upward slanting solid blue lines of decreasing slope as a function of increasing wavelength/filter color). Each Cepheid's luminosity deviation has the same sign, and each appears to have a very similar magnitude in each of the plots, independent of the wavelength being considered. That is, these coherent deviations are largely achromatic, a signature of them being distance modulus errors shared in an equivalent way in each plot, regardless of the slope of the individual lines as a function of wavelength.

The Gaia parallaxes as currently understood, preliminarily corrected and currently published, cannot be totally error free. Indeed, their statistical uncertainties are published and readily available. We therefore equate the bandpass-averaged offsets derived from the sub-panels in Figure 12 to be valid {\it differential} corrections to the Gaia parallaxes analyzed by Breuval et al. (2021).
We note also that if the Cepheid parallaxes as a whole are not representative of the greater Gaia database then there is still the possibility of another global zero-point shift, which this restricted-sample MW recalibration is incapable of revealing; however, see the following sections for one external test for that systematic.

The scatter in the correlations of magnitude and color seen in the sub-panels of Figure 13 
are, respectively,  $\sigma_V = \pm 0.076, \sigma_I = \pm 0.023, \sigma_J = 0.031, \sigma_H = \pm 0.034, \sigma_K = \pm 0.039$ mag. 
The PLC formalism is thereby delivering distances to individual Cepheids in this Milky Way sample at the $\pm$ 1-4\% level per star.

It has to be emphasized that, for any given Cepheid, {\it only one differential distance modulus correction is being applied} to each PL relation, independent of wavelength. As such, the relative impact of this one correction is larger for the long-wavelength PL relations whose intrinsic widths are smaller than the intrinsic widths seen at shorter wavelengths. So the questions become: What fraction of the scatter in the residual-residual plots is removed by this single offset being applied to each of the frames? And then, from a different perspective:  What, if anything, do these star-by-star offsets do to the run in the total width (i.e., scatter) in the resulting PL relations as a function of increasing wavelength? Figure 14 clearly addresses the second question; while the answer to the reduced scatter is:  up to a factor of 4 improvement is seen at the longest wavelengths.

\begin{figure*} 
\includegraphics[width=18.0cm, angle=-0]{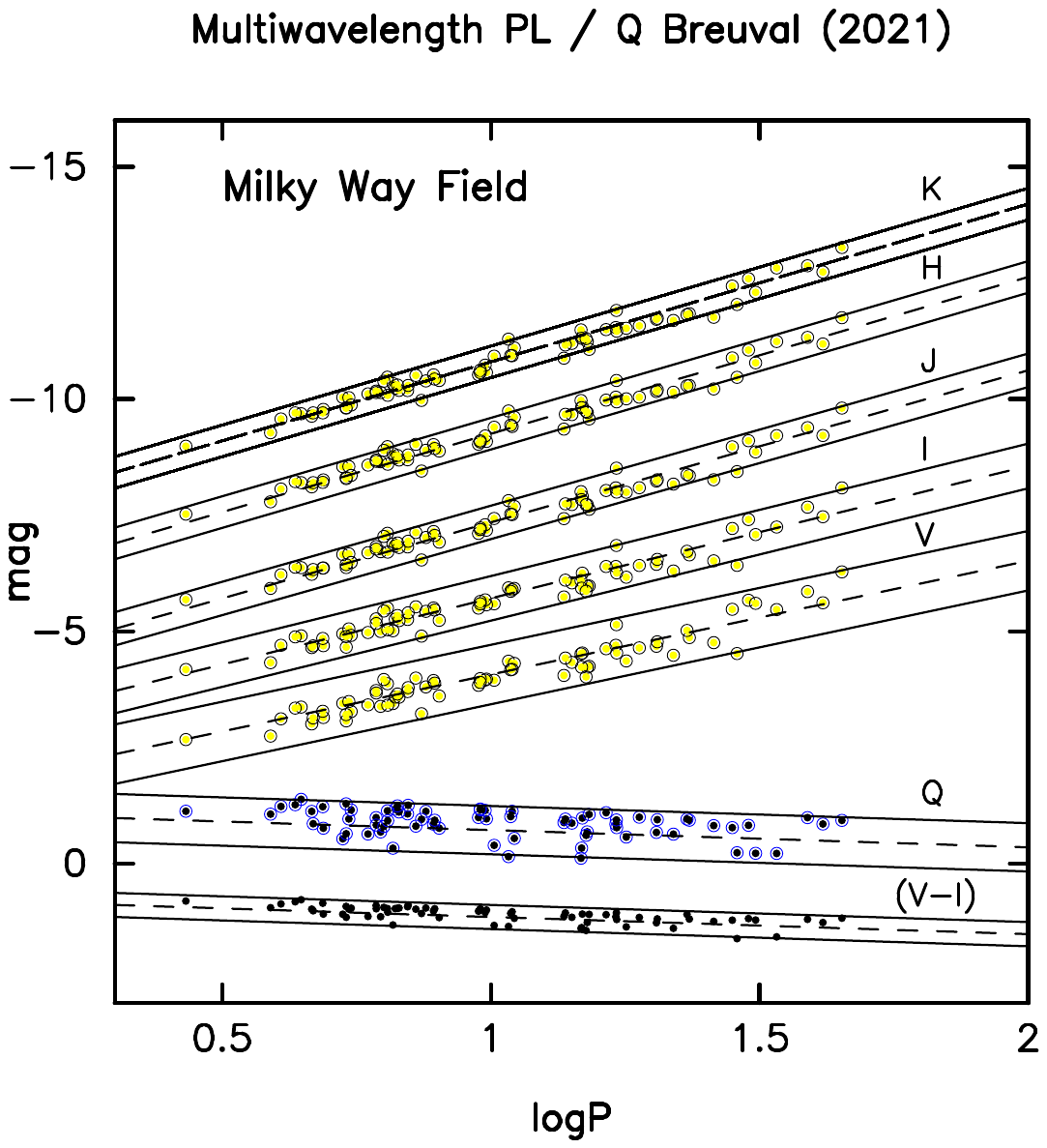} 
\caption{Milky Way intrinsic period-luminosity (VIJHK, yellow points, above) and period-color relations (Q, blue points and (V-I), black points, below) as taken from Breuval et al. (2021). Individual relations are offset for clarity by forcing reduced overlap.
}
\end{figure*}

\begin{figure*} 
\includegraphics[width=18.0cm, angle=-0]{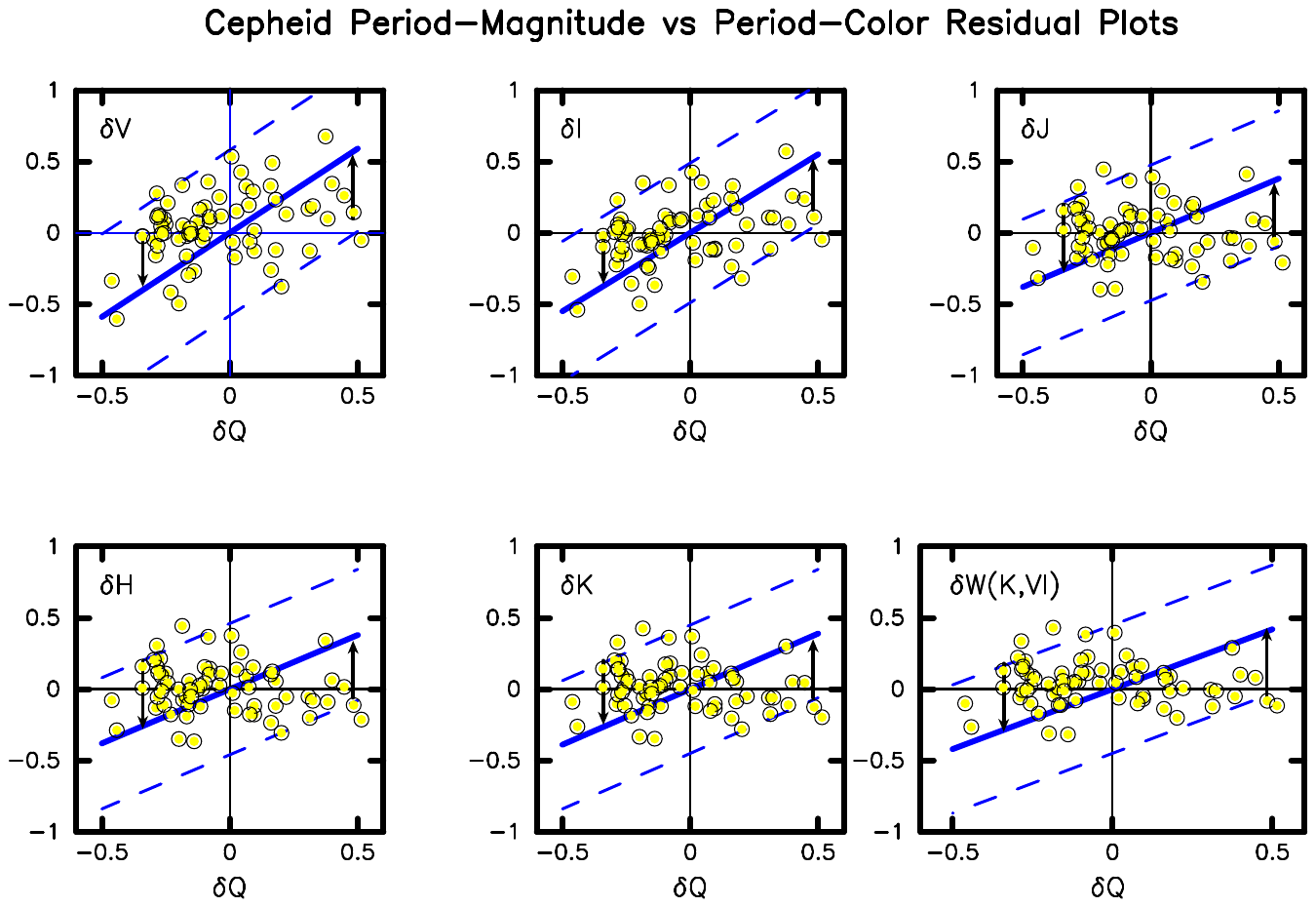} 
\caption{Milky Way magnitude and color deviations from the period-luminosity and the reddening-free (Q, Qualine) period-color relations  as a function of filter wavelength (VIJHK and W) shown in separate panels from left to right and top to bottom. The intrinsic correlation between pairs of luminosity and color residuals are shown as upward-slanting blue, paralleled by two-sigma (broken) lines bounding the observations. The same random sampling of data points are  shown in red in each of the panels in order to ease the positional inter-comparison of Cepheids from plot to plot. The remaining data points are shown as small black points. Finally, two red points have black arrows going vertically from the data point to the intrinsic blue line.  Obvious to the eye, these vectors are of the same sign and closely of the same magnitude independent of the wavelength of the filter/plot being examined. We attribute these residuals to errors in the Gaia parallaxes, which are achromatic, by their very nature. 
}
\end{figure*}

\begin{figure*} 
\includegraphics[width=18.0cm, angle=-0]{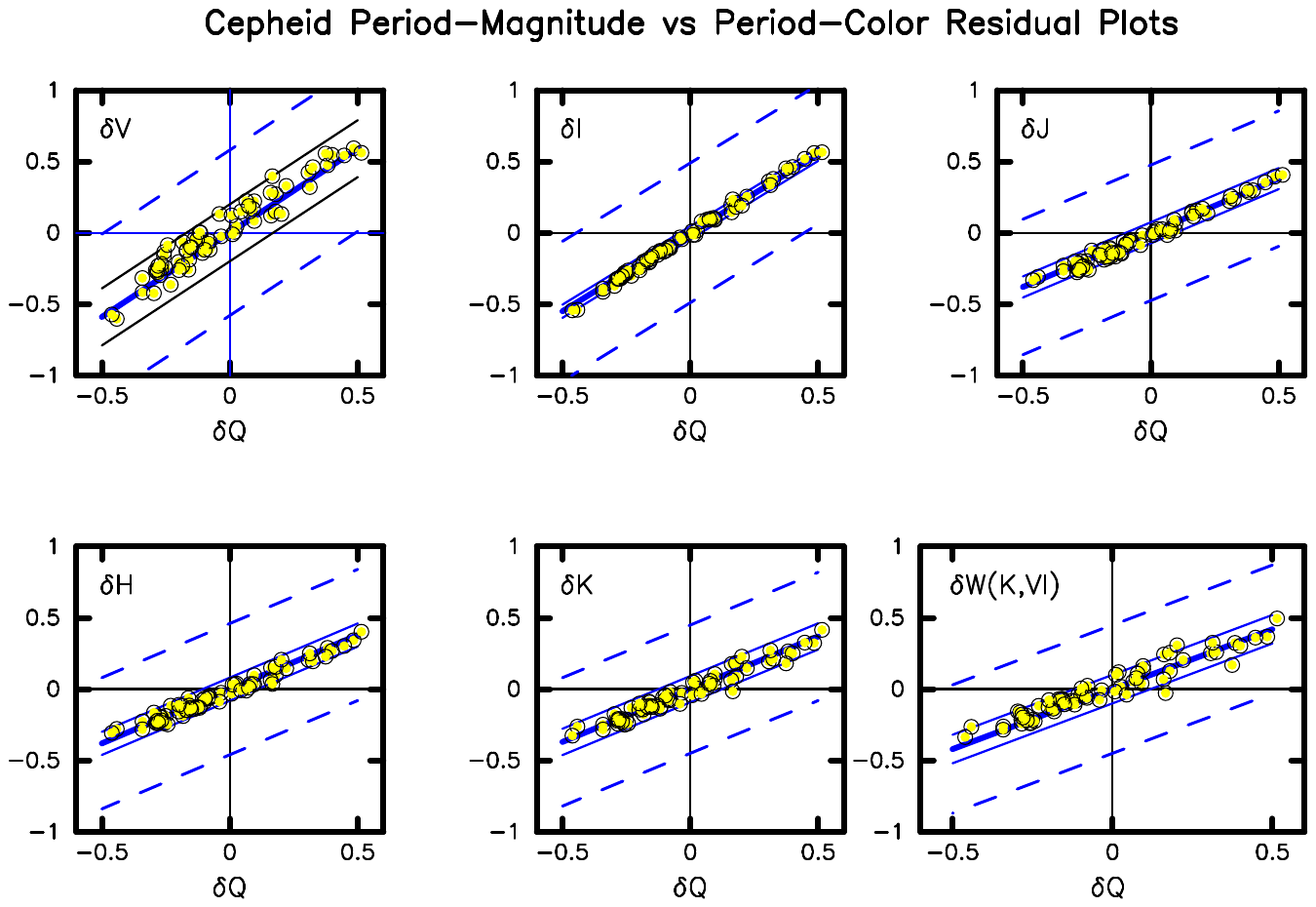} 
\caption{Milky Way parallax-corrected magnitude and color deviations from the period-luminosity relation and the reddening-free (Q, Qualine) period-color relations (from Figure 1) as a function of filter wavelength (VIJHK and W) shown in separate panels from left to right and top to bottom. The intrinsic correlation between pairs of luminosity and color residuals are shown as upward-slanting blue, paralleled by two-sigma (broken) lines bounding the observations (as taken from Figure 1, to emphasize the order-of-magnitude decrease in the scatter of the observations around the intrinsic blue lines.  
}
\end{figure*}

\begin{figure*} 
\includegraphics[width=18.0cm, angle=-0]{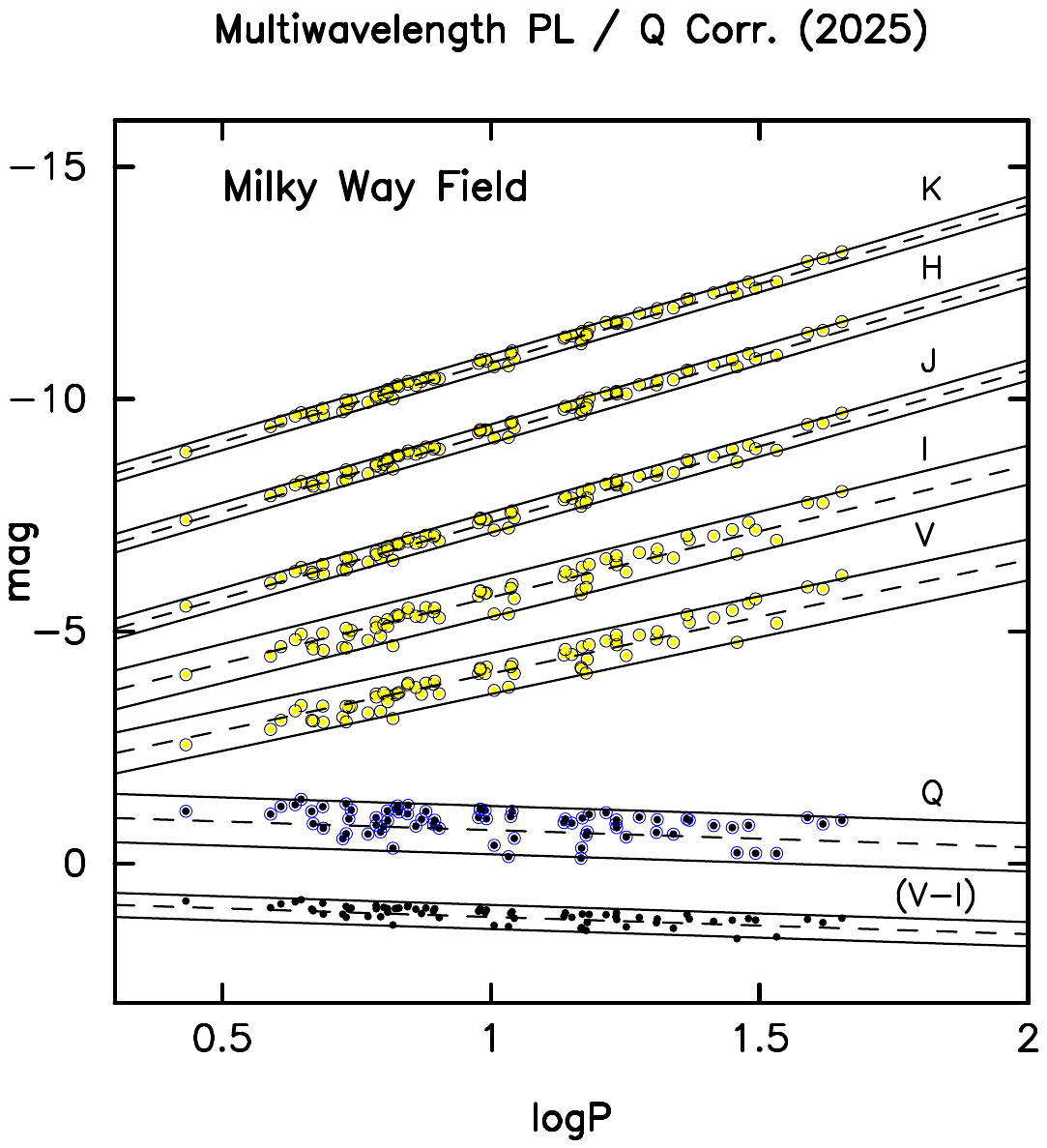} 
\caption{Gaia-corrected intrinsic period-luminosity (VIJHK, yellow points, above) and period-color relations (Q, blue points and (V-I), black points, below) as taken from Breuval et al. (2021). Individual relations are again offset for clarity by forcing reduced overlap. Note the smooth and monotonic decrease of the widths of the individual PL relations as a function of increasing wavelength V through K.
See text for details and implications.
}
\end{figure*}
\begin{figure*} 
\includegraphics[width=18.0cm, angle=-0]{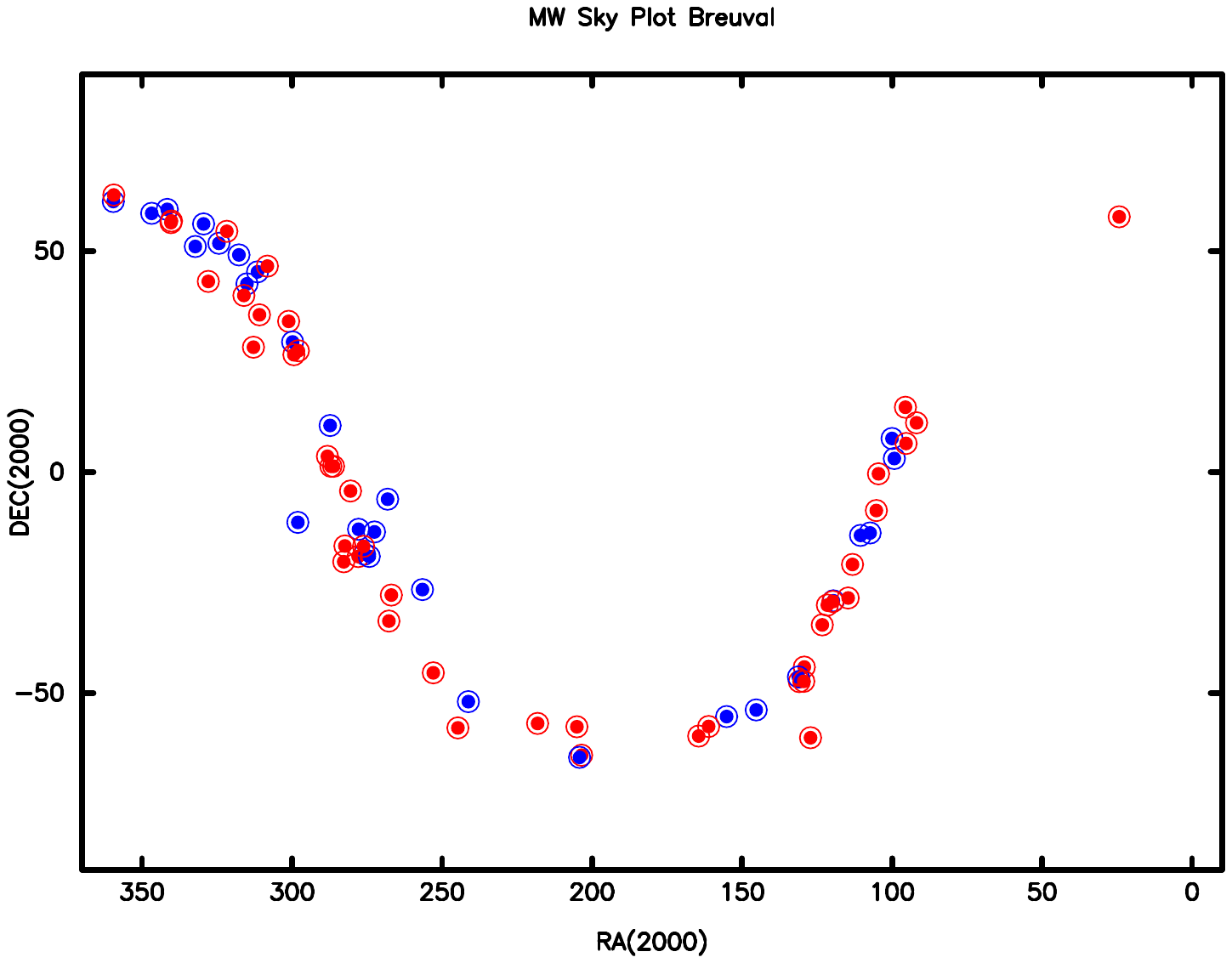} 
\caption{All-sky distribution of Milky Way Cepheids. As is obvious, and expected, most of the Cepheids in this sample are found in the plane of the MW. The points are color-coded by the sign of their individual parallax corrections, in search of spatial dependencies. With the possible exception of an excess of red points from 0 to 130 deg in RA, no obvious correlations within this highly restricted sample are found.
}
\end{figure*}
\begin{figure*} 
\includegraphics[width=10.0cm, angle=-0]{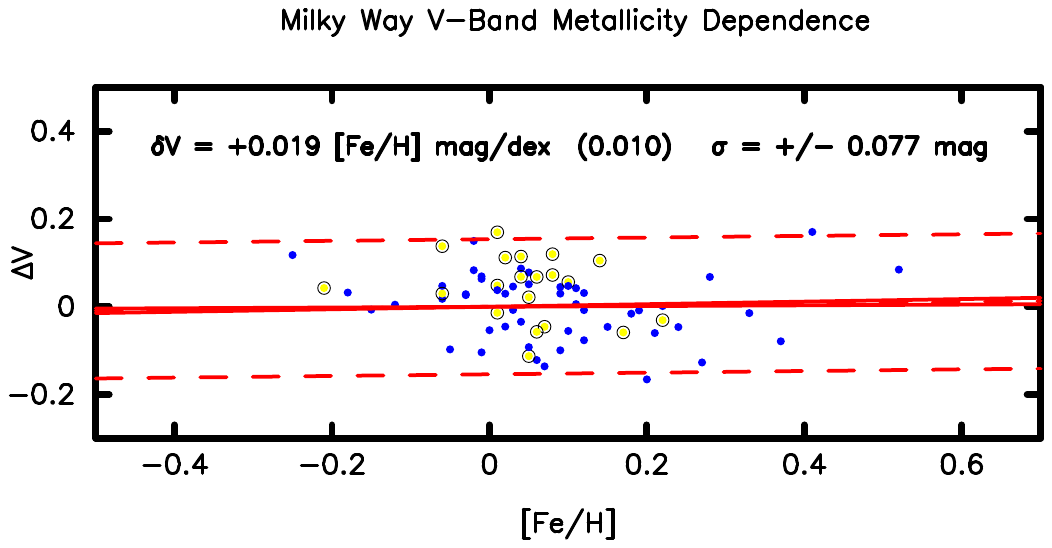} \includegraphics[width=10.0cm, angle=-0]{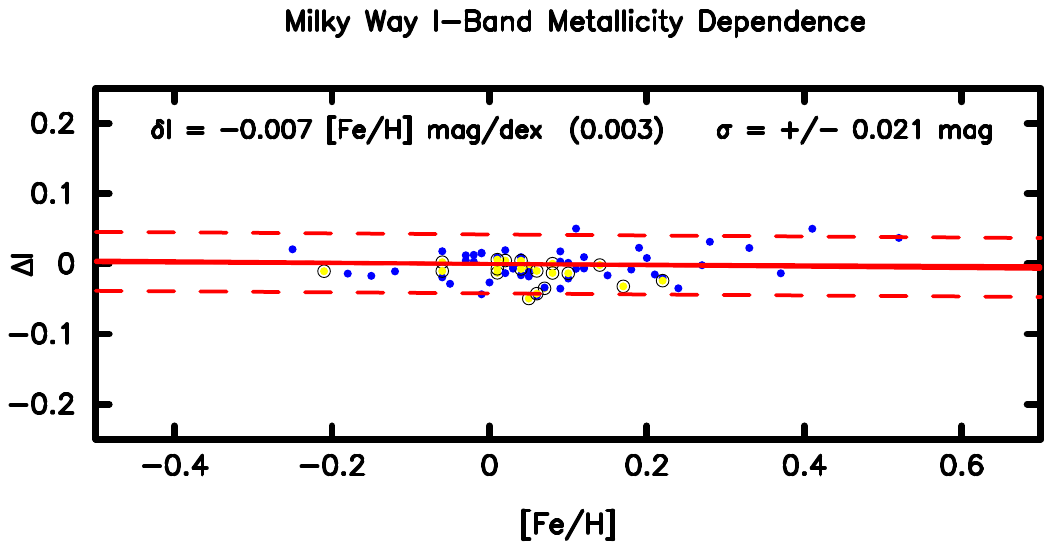} 
\includegraphics[width=10.0cm, angle=-0]{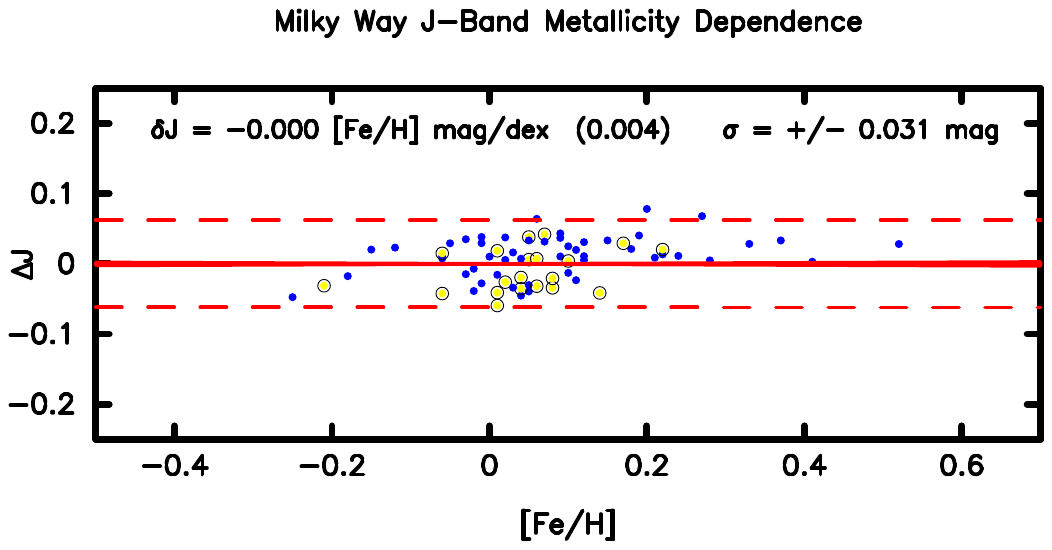} \includegraphics[width=10.0cm, angle=-0]{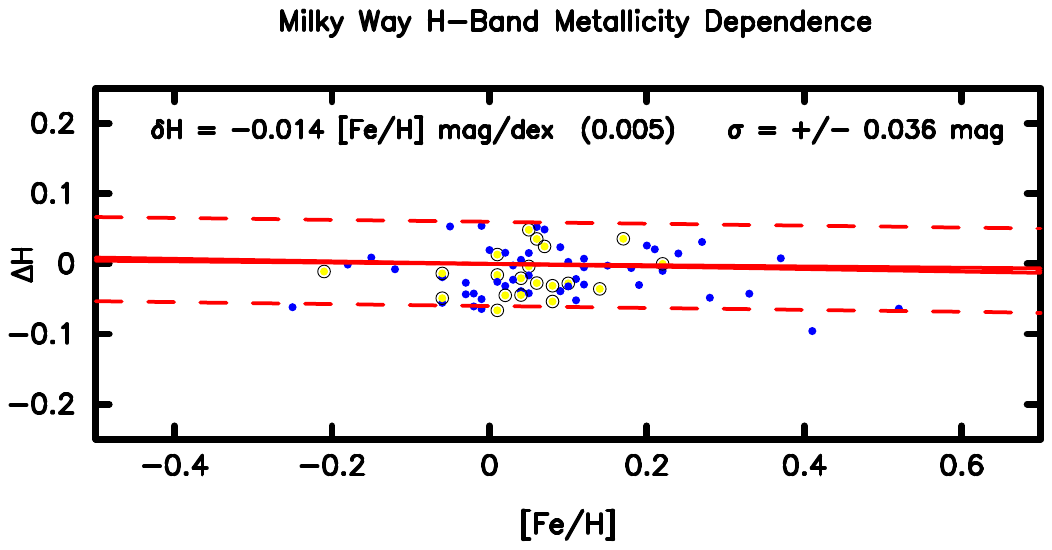} 
\includegraphics[width=10.0cm, angle=-0]{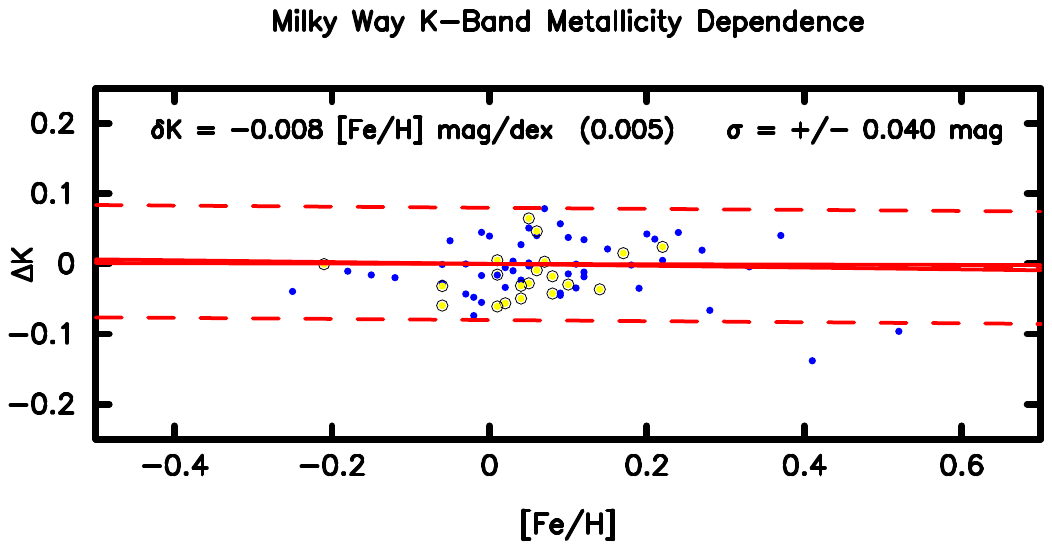} 
\caption{Correlation of final magnitude residuals in the optical (VI) and NIR (JHK) with metallicity, [Fe/H]. In no case is there any statistically detectable relationship between the two variables, with the scatter around zero being at or below $\pm$0.04 mag, in all but the V band which has a scatter of $\pm$ 0.077 mag. The reason for the larger scatter at V remains unexplained at this time.  
}
\end{figure*}
\clearpage
\subsection{Wavelength-Dependent Metallicity Sensitivities}
{ Spectroscopic metallicities have been measured for Milky Way Cepheids and span a range of -0.2 to 0.5 dex in [Fe/H], with typical uncertainties of $\pm$0.12 dex (Luck \& Lambert (2011) and Luck 2018). Lesmasleet al. (2018) and Andrievsky \& Koutyukh (2011) estimate the Milky Way Cepheid [Fe/H] metallicity gradient amounts to -0.055 dex/kpc. This gradient is consistent with the independently measured gas-phase [O/H] HII region abundances, which have a gradient of -0.044 dex/kpc (e.g., Menestre-Delgado et al. 2022). The abundance range for Milky Way Cepheids is not large, but it does span nearly one third of the total extragalactic range, and has proven sufficient to allow the Milky Way radial gradient in abundance to be measured; thus, significant wavelength dependencies as searched for above, should also be measurable. Those diagnostic plots are shown in the five sub-panels of Figure 16.}

The unweighted fits to the data 
given across the top of each plot. The slopes range from +0.019 mag/dex in the V band to -0.014 mag/dex in the I band, where the data cover a small metallicity range of 0.7 dex in [Fe/H]. The scatter in V is the largest, $\pm$ 0.077~mag, while the scatter in the other four bands is significantly smaller, falling between $\pm$0.040 and 0.021~mag. 

\section{Step 4: Nearby Cepheids (HST) and Cepheids in Open Clusters  (Gaia)}
\subsection{Step 4a: Two Samples Combined}
In an earlier paper Breuval et al. (2020) considered a subset of nearby Cepheids with high-precision HST parallaxes, combined with Cepheids in open clusters where parallaxes were available for a significant number of main sequence stars whose average parallax could be applied to the Cepheids. The sample was approximately evenly split between the 9 HST parallaxes and the 7 Gaia parallaxes, equitably spreading the risks between two systems, with each focusing either on  near, or far samples of field and cluster Cepheids, respectively. In this first sub-section we look for parallax corrections around the mean of the parallaxes obtained for the combined sample.

Multi-wavelength VIJHK PL relations for the combined sample of Milky Way Cepheids observed by HST in the nearby field (9 blue points), and those in open clusters observed by Gaia (7 red points) are shown in Figure 17. Clearly, there is a systematic offset, at all wavelengths, between the two samples. This dichotomy was noted in Madore \& Freedman (2025), and the consequences of one approach to resolving that issue were discussed in that paper. We undertake a similar analysis here, using the discrepancy as a currently available means of testing the Gaia parallaxes for a true zero point offset with respect to an independent (HST) zero point. This first step turns a blind eye on the dichotomy which will presumably be seen in the residual-residual plots as due to achromatic magnitude residual around the average. The Q and $(V-I)_o$ period-color relations at the bottom of Figure 17 suggest that the dichotomy in the PL plots is not shared in these distance-independent period-color plots, which should be responsible for the deterministic magnitude scatter in each of the PL relations.

\begin{figure*} 
\includegraphics[width=18.0cm, angle=-0]{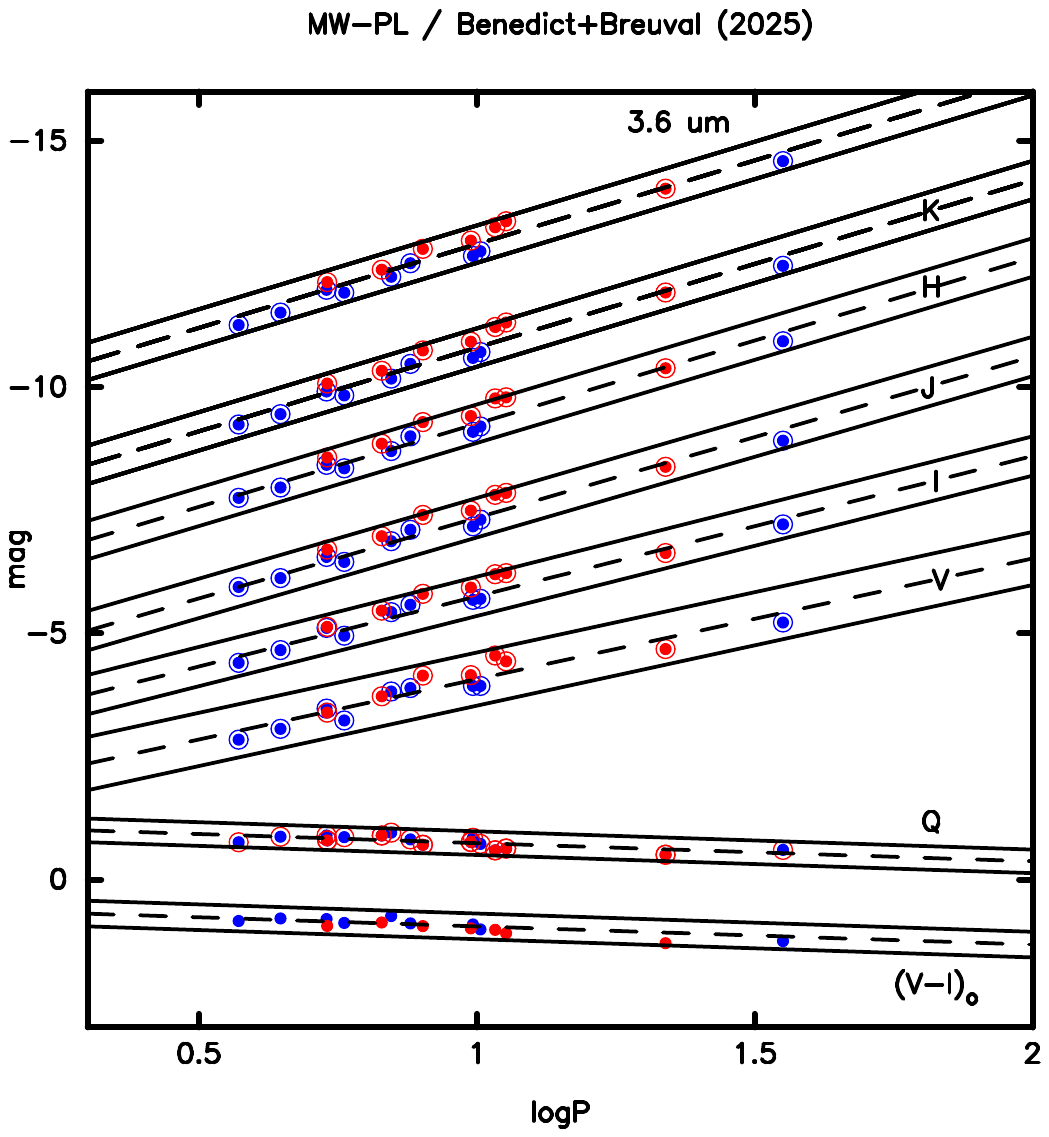} 
\caption{Multi-wavelength VIJHK PL relations for the combined sample of Milky Way Cepheids observed by HST in the nearby field (9 blue points), and those in open clusters observed by Gaia (7 red points). Clearly there is a systematic offset, at all wavelengths, between the two samples.}
\end{figure*}

The magnitude-residual versus color-residual plots are shown in Figure 18 for all 5 of the PL relations and for the reddening-free Wesenheit (K,VI) data. Again, the HST and Gaia data are distinguished by the same two colors as in Figure 17. The HST (blue points) data are systematically offset, and fainter than the red Gaia cluster data points, but all of the deflections away from the expected fiducial lines (plotted as solid, upward sloping blue lines) are strongly cross-correlated in magnitude and in sign from plot to plot. Those cross-correlations are the signature of distance moduli errors, whether they arise in one sample or across both samples. 

For each Cepheid we calculate the average magnitude offset and apply that single number to the original multi-wavelength data, and plot the results in Figure 19. The change between the PL relations, from before to after the parallax correction, is obvious and striking. The scatter has systematically decreased across all bandpasses, and the internal scatter from band pass to band pass decreases monotonically but more strongly than in the uncorrected PL plots as a function of increasing wavelength.

Figure 20 shows the residual-residual plots with the data corrected for distance moduli errors. The resulting scatter can be attributed to the propagated photometric errors in the input apparent magnitudes and colors of the Cepheids drawn from photometric heterogeneous surveys.

Without suggesting whether one parallax method is better than the other, this application of our methodology splits the difference, but ultimately it is insensitive to underlying systematics.

We now apply the same methodology to each of the parallax sub-samples individually, in the two sub-sections that follow. A quantification of the zero-point differences between the HST and GAIA parallaxes will then be made obvious, with tentative conclusions being given at the end of this section.

\begin{figure*} 
\includegraphics[width=18.0cm, angle=-0]{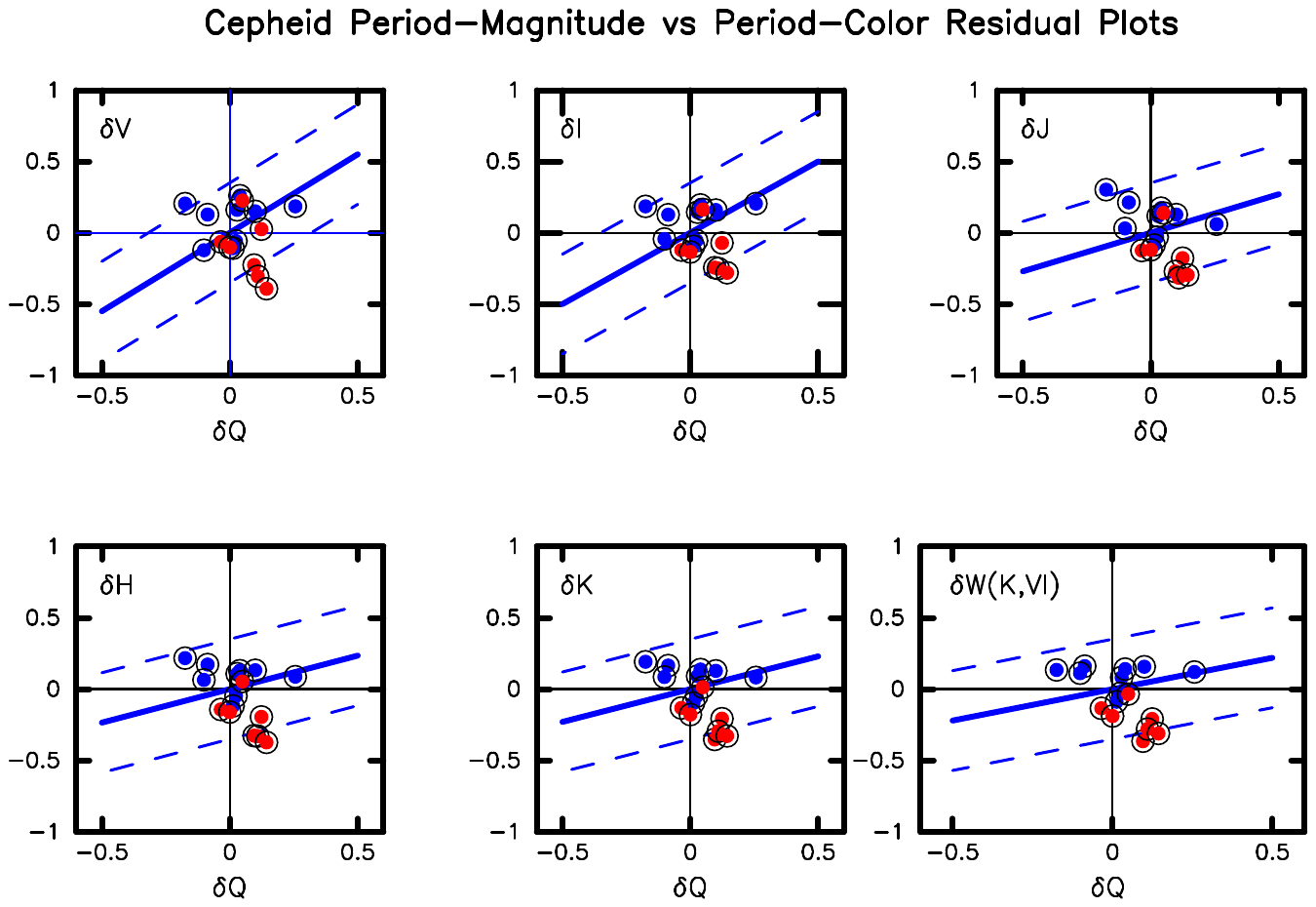} 
\caption{Same as Figure 10, except considering the Breuval et al. (2020) HST (blue) and Gaia (red) data individually distinguished.  }
\end{figure*}

\begin{figure*} 
\includegraphics[width=18.0cm, angle=-0]{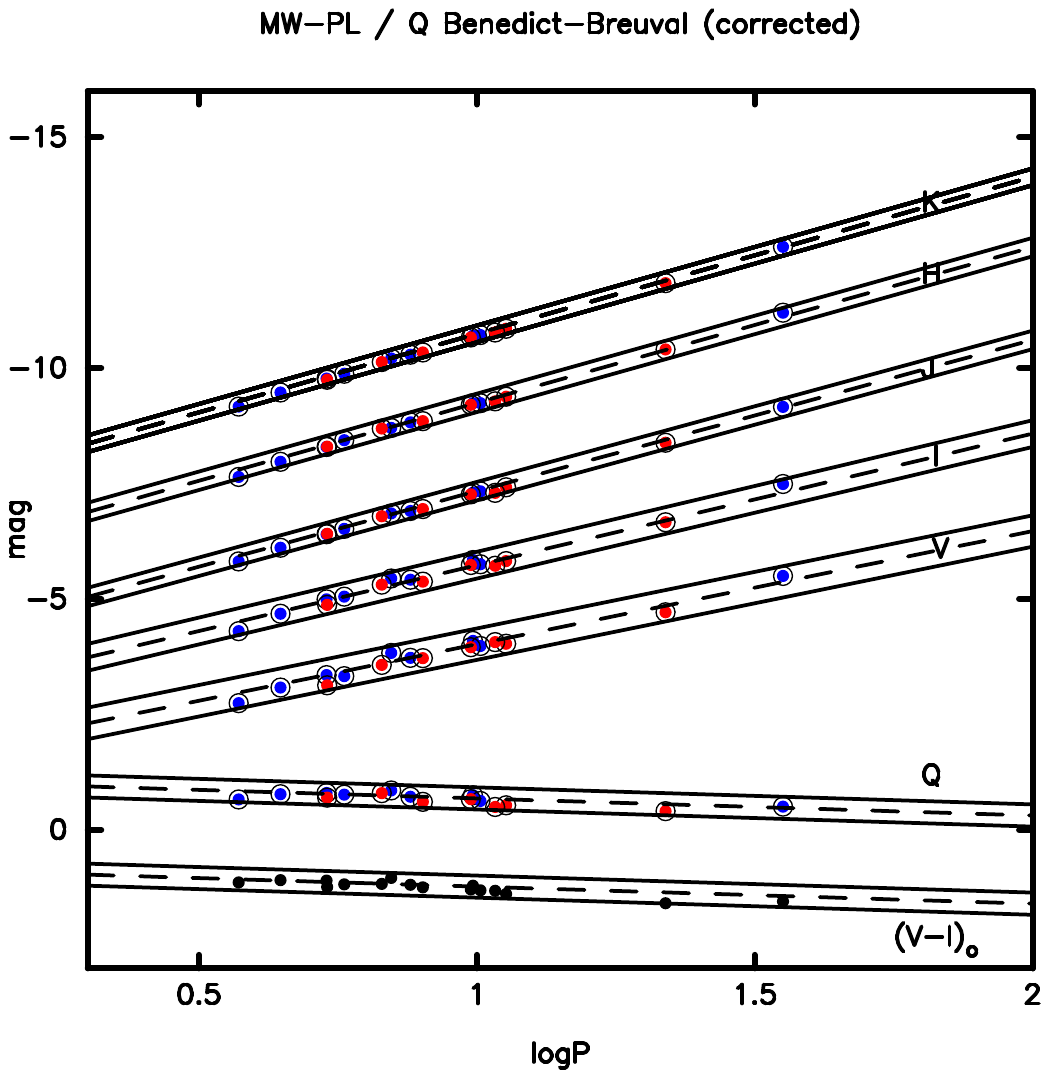} 
\caption{Same as Figure 17, except the data plotted here have been corrected for parallax offsets. }
\end{figure*}

\begin{figure*} 
\includegraphics[width=18.0cm, angle=-0]{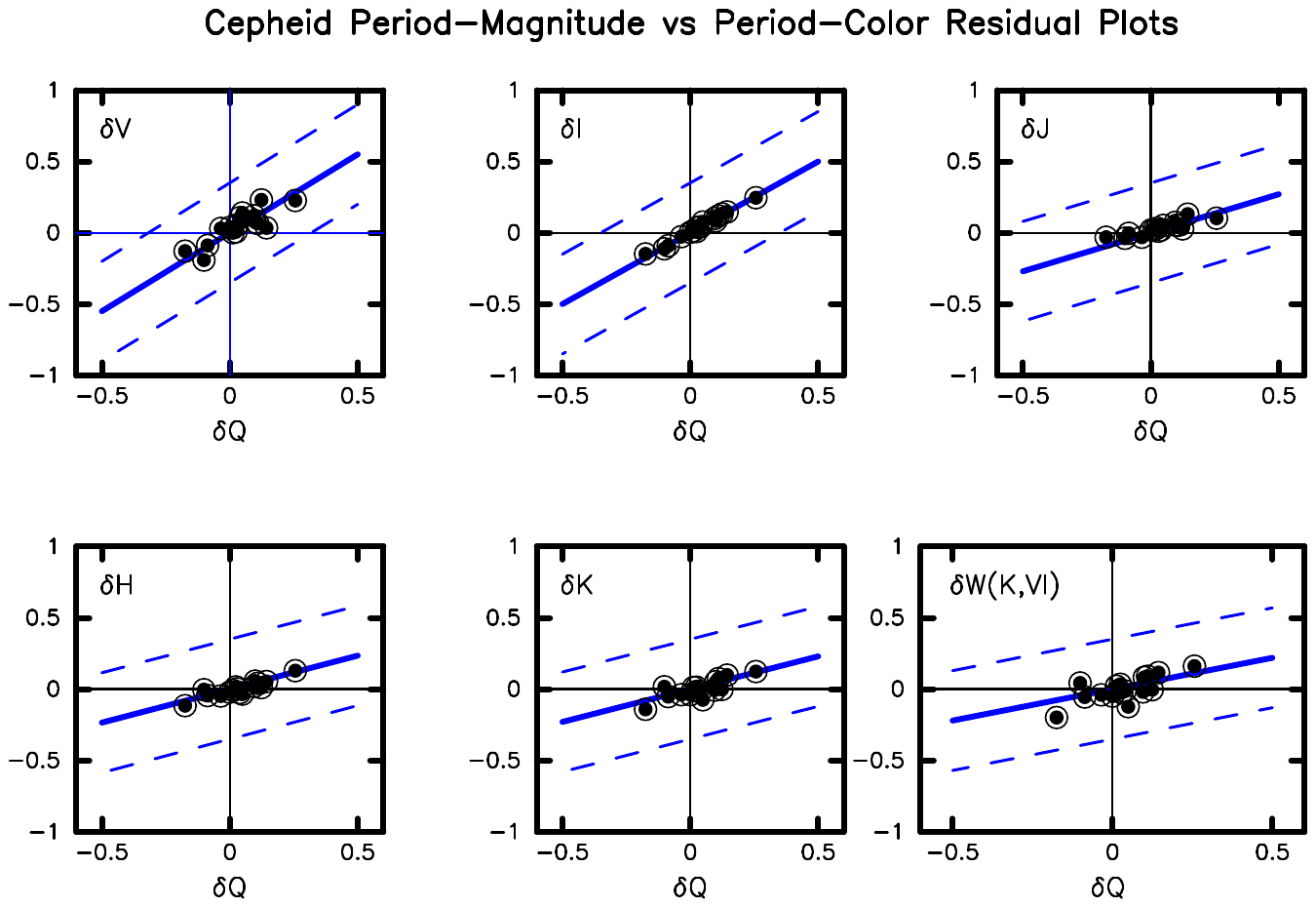} 
\caption{Same as Figure 18 except the residuals  in magnitude have been corrected for the Gaia parallax offsets, as described in the text.}
\end{figure*}

\subsection{Step 4b: Gaia Parallaxes for Cepheids in MW Clusters Alone}

Breuval et al (2020)  and others (e.g., Table 1 in Madore \& Freedman 2025) have made the case that higher-precision parallaxes for Cepheids  that are members of open clusters can be had from averaging over the significantly larger population of main sequence stars that are also members of the same clusters. We take that subset of Cepheids and their parallaxes, as tabulated by Breuval et al. and apply our parallax updating methodology used above on the combined data set.

Figure 21 shows the VIJHK PL relations for cluster Cepheids alone. At the bottom of the panels are the two period-color relations based on $(V-I)_o$ and Q intrinsic colors. The dashed lines are a fit to the data. However the flanking (solid) lines are the two-sigma boundaries carried over from Section 6.1. It is immediately clear that this Gaia-selected subset of Cepheids do not fill any of the PL relations, and that the points have a surprisingly low dispersion.

Progressing to the residual-residual plots given in Figure 22, we see that while the points are not inconsistent with the expected intrinsic (solid blue) lines, there is still however a highly significant cross-correlation of (vertical) magnitude-residual deflections across the bands. Averaging the parallax errors and subtracting them on a star-by-star basis results in the parallax-corrected PL plots shown in Figure 25 and the corrected residual-residual plots in Figure 24.


\begin{figure*} 
\includegraphics[width=18.0cm, angle=-0]{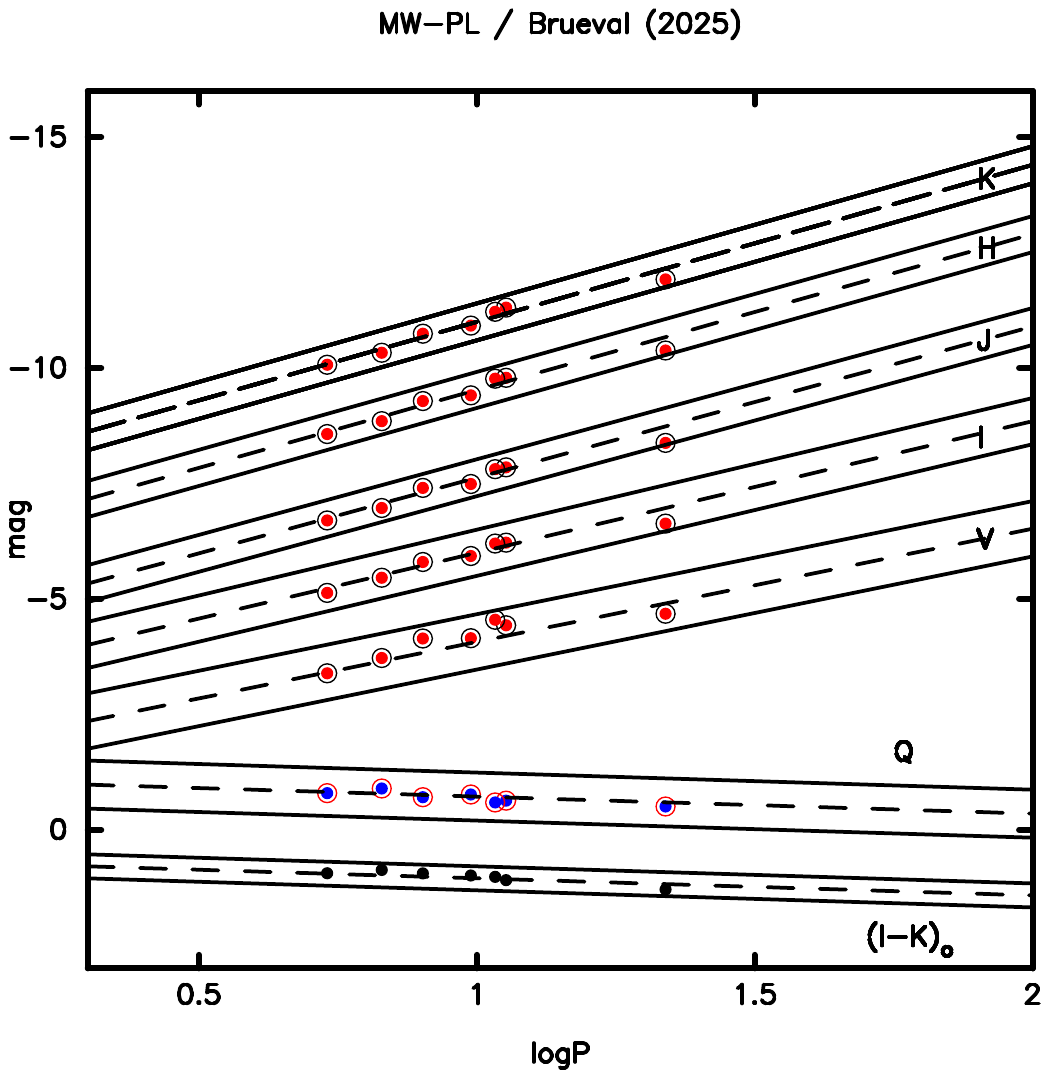} 
\caption{ Same as Figure 17, except that only the Gaia-parallax data from Breuval et al. (2020) are shown.
}
\end{figure*}
\begin{figure*} 
\includegraphics[width=18.0cm, angle=-0]{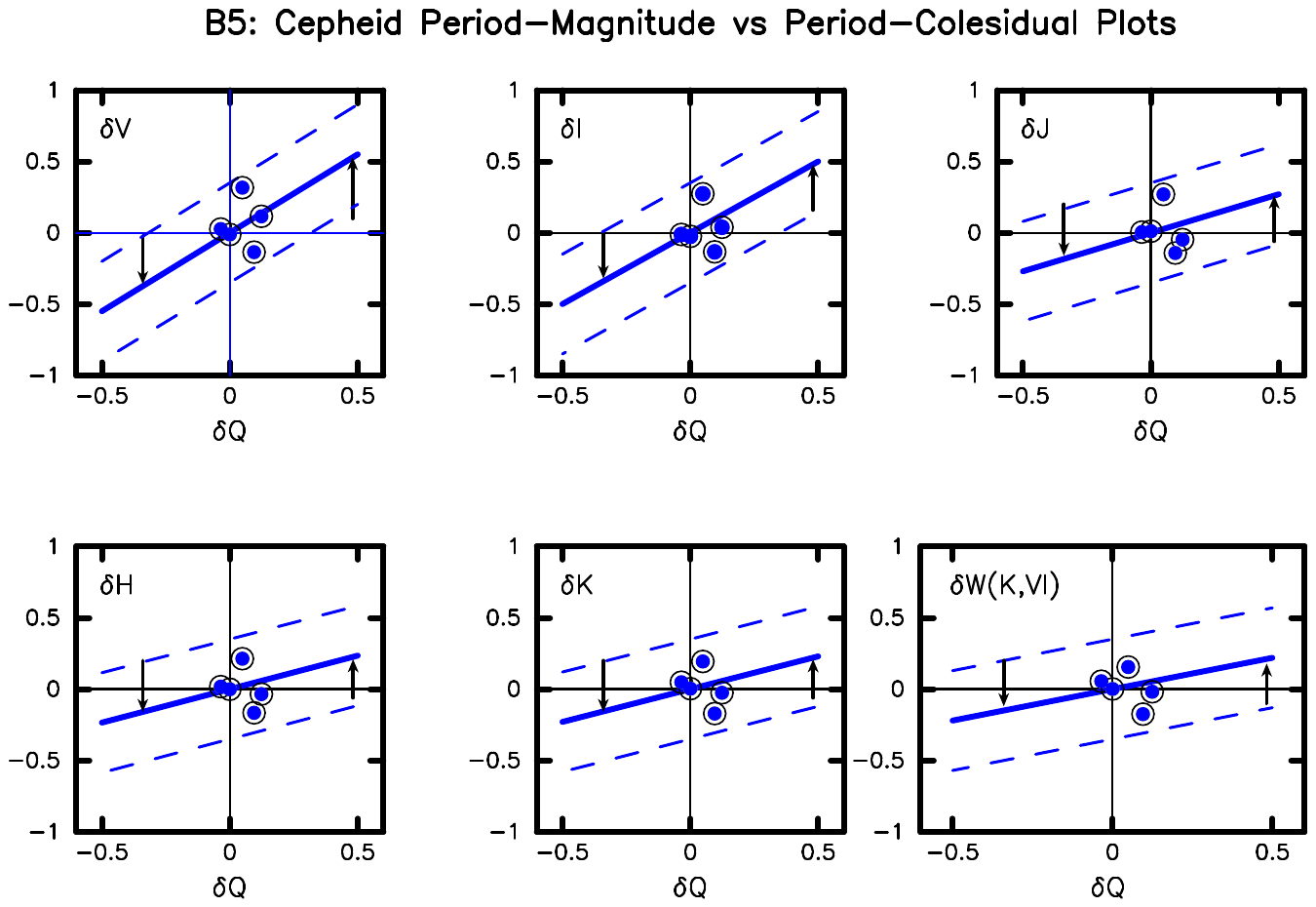} 
\caption{Same as Figure 18, except that only the Gaia-parallax data from Breuval et al. (2020) are shown.
}
\end{figure*}

\clearpage
\begin{figure*} 
\includegraphics[width=18.0cm, angle=-0]{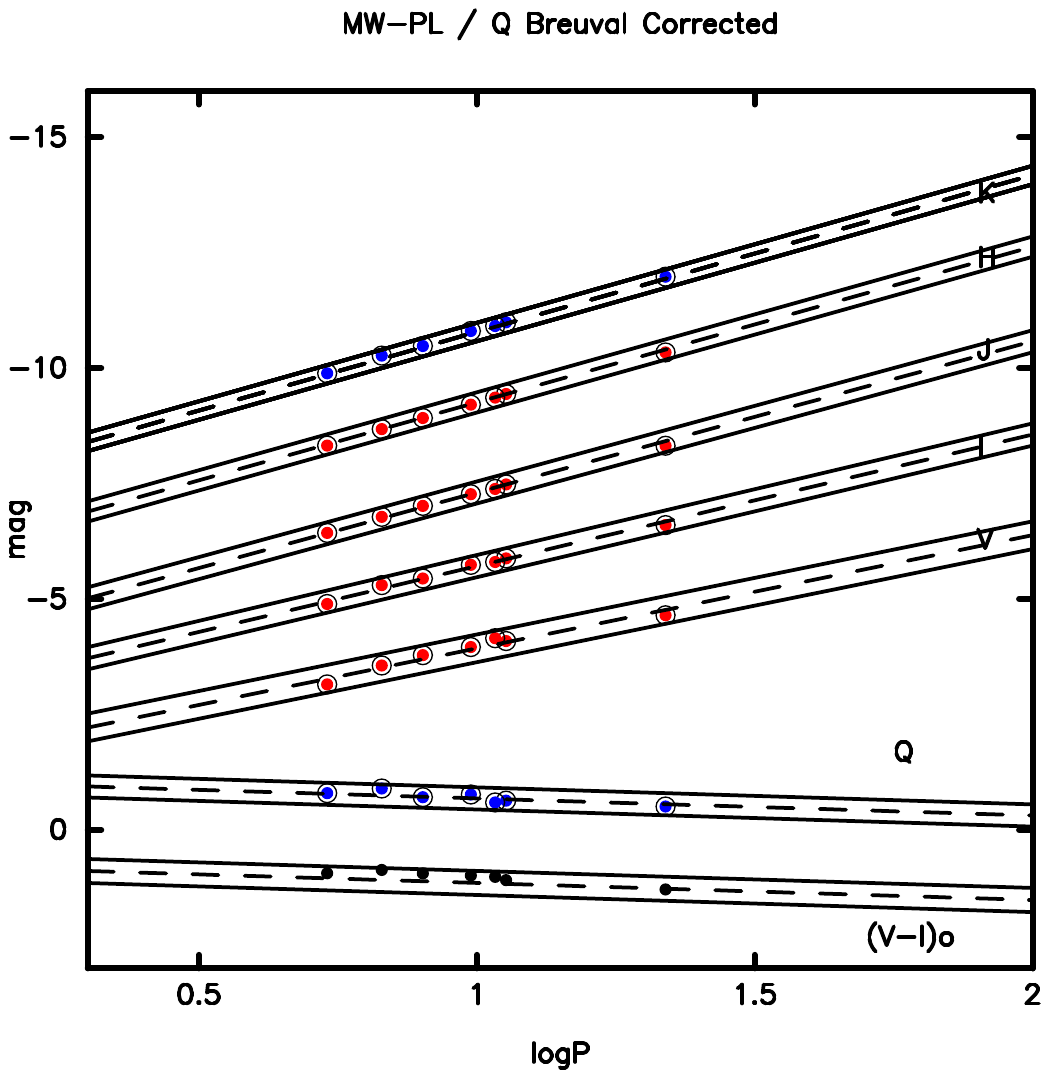} 
\caption{Same as Figure 19, except that only the Gaia-parallax data from Breuval et al. (2020) are shown.
}
\end{figure*}

\begin{figure*} 
\includegraphics[width=18.0cm, angle=-0]{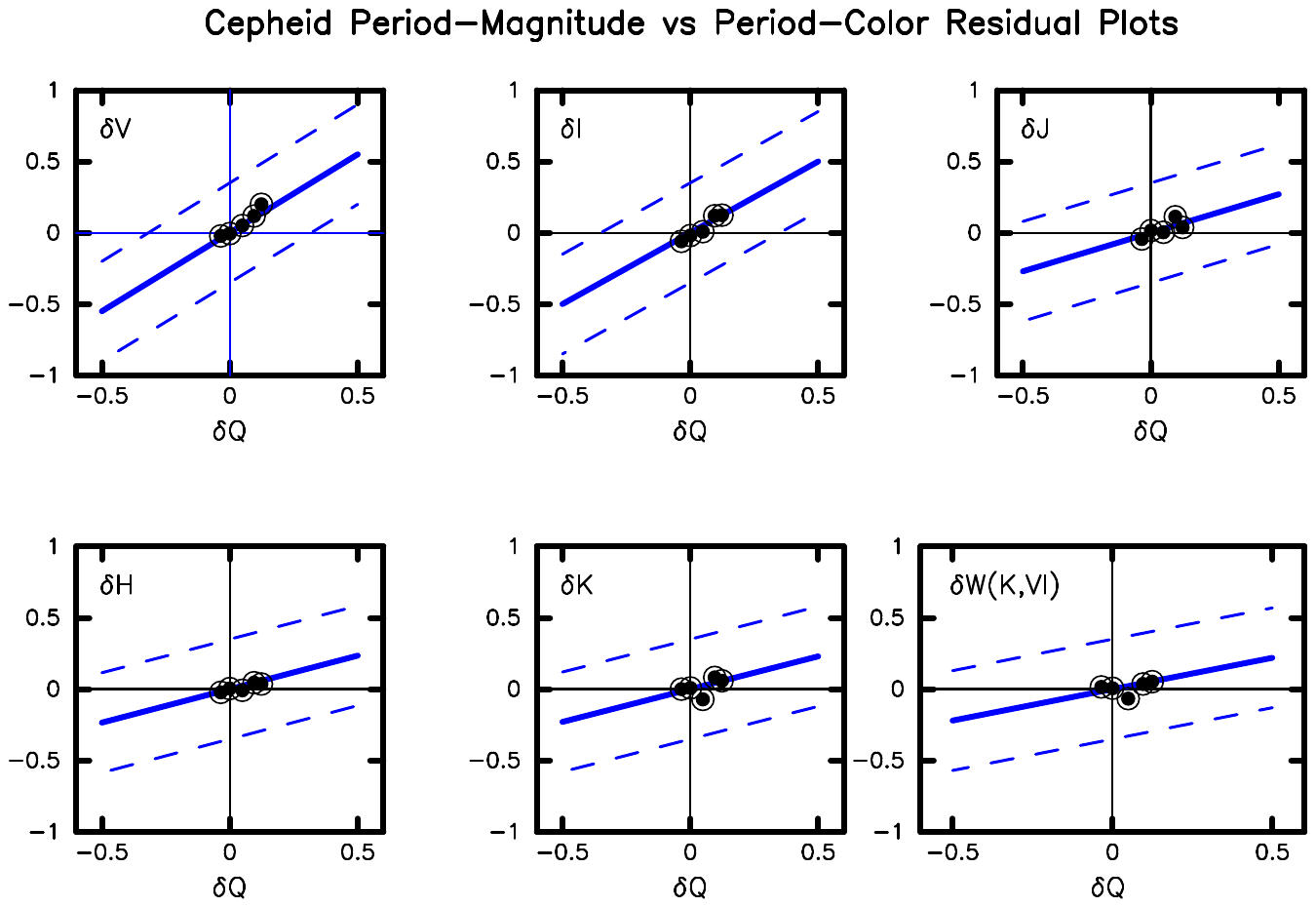} 
\caption{Same as Figure 20, except that only the Gaia-parallax data from Breuval et al. (2020) are shown.
}
\end{figure*}

\clearpage

\subsection{Step 4c: HST Parallaxes for Field Cepheids in Alone}

We now examine the Benedict et al. (2007) subsample of nearby Milky Way Cepheids observed on the HST (Fine-Guidance Sensor) zero point.

Figure 25 shows the absolute-magnitude PL relations for VIJH \& K (red points) as published.
The reddening-free Q color index plotted as a function of period is shown as blue points towards the bottom of the panel, followed by the individually reddening-corrected $(V-I)_o$ data points as a function of period. All of these relations show larger widths than the Breuval et al. (2020) subsample, possibly due to there being more stars in this sample.

Figure 26 shows the magnitude-residual versus color-residual plots for the Benedict et al. Cepheids. The expected trends are not seen in the data; however, there is evidence for some achromatic correlated scatter between the panels.
Averaging the achromatic terms for each star, and subtracting the parallax correction from the absolute magnitudes results in Figure 27 where we note the strong visual correlation of the positions of individual stars in the period-color relations with the magnitude positions in each of the PL relations above them. We also note that the parallax corrections have preserved the monotonically increasing widths of the PL relations from the NIR to the optical.

Finally, we show in Figure 28 the parallax-corrected residual-residual plots, where the data now conform to the intrinsic (solid blue) lines.
\begin{figure*} 
\includegraphics[width=18.0cm, angle=-0]{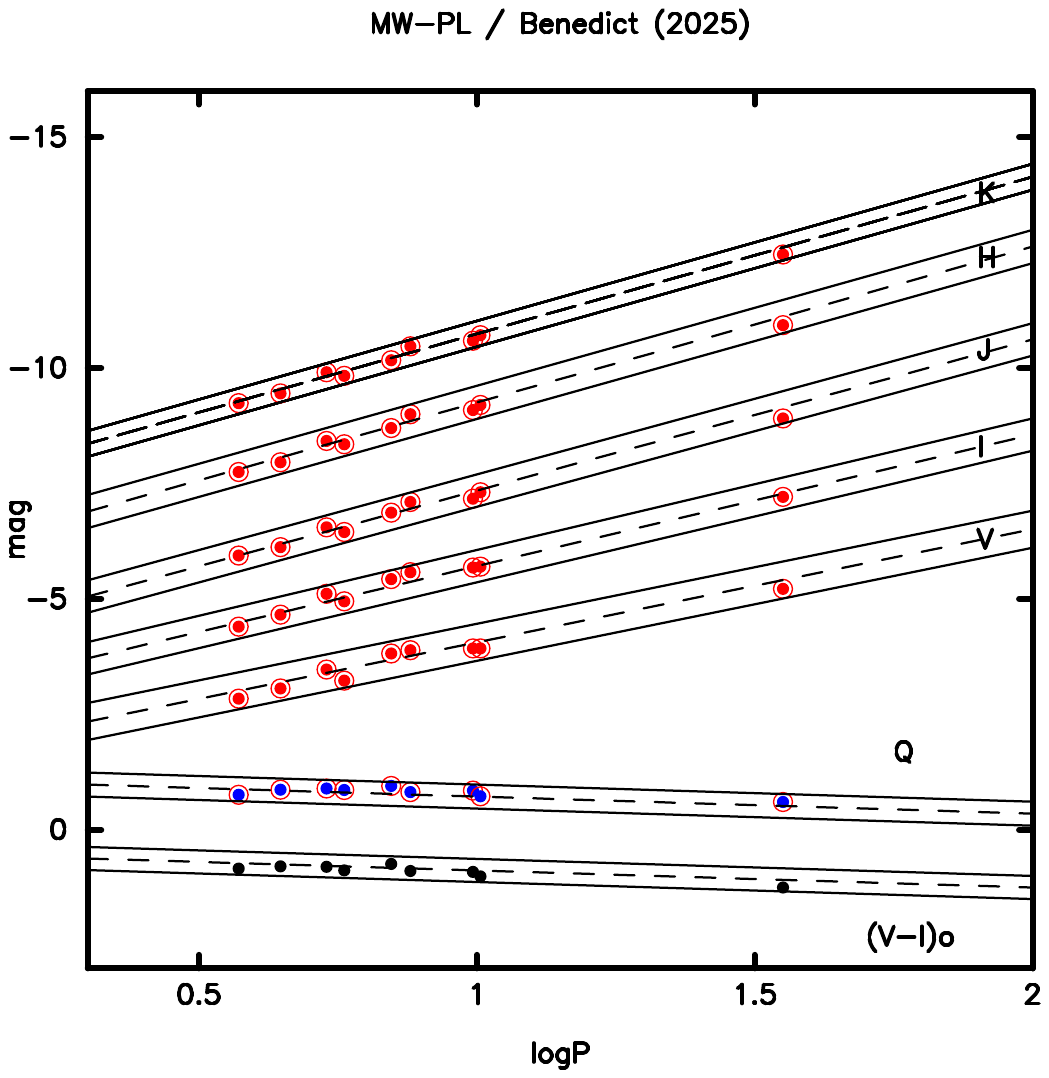} 
\caption{Same as Figure 19, except that only the HST-parallax data from Benedict et al. (2007) are shown.
}
\end{figure*}

\begin{figure*} 
\includegraphics[width=18.0cm, angle=-0]{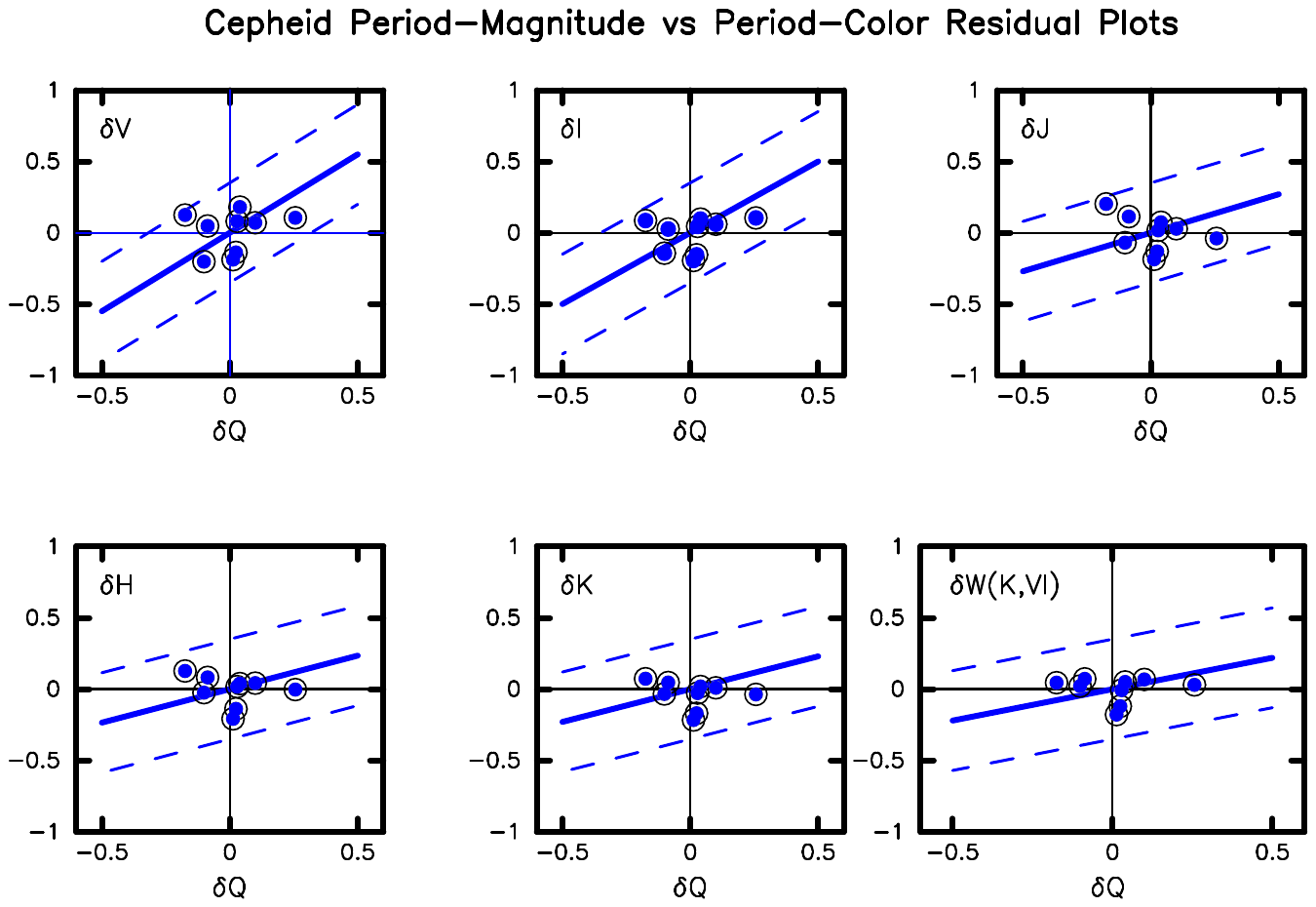} 
\caption{Same as Figure 20, except that only the HST-parallax data from Benedict et al. (2007) are shown.
}
\end{figure*}

\begin{figure*} 
\includegraphics[width=18.0cm, angle=-0]{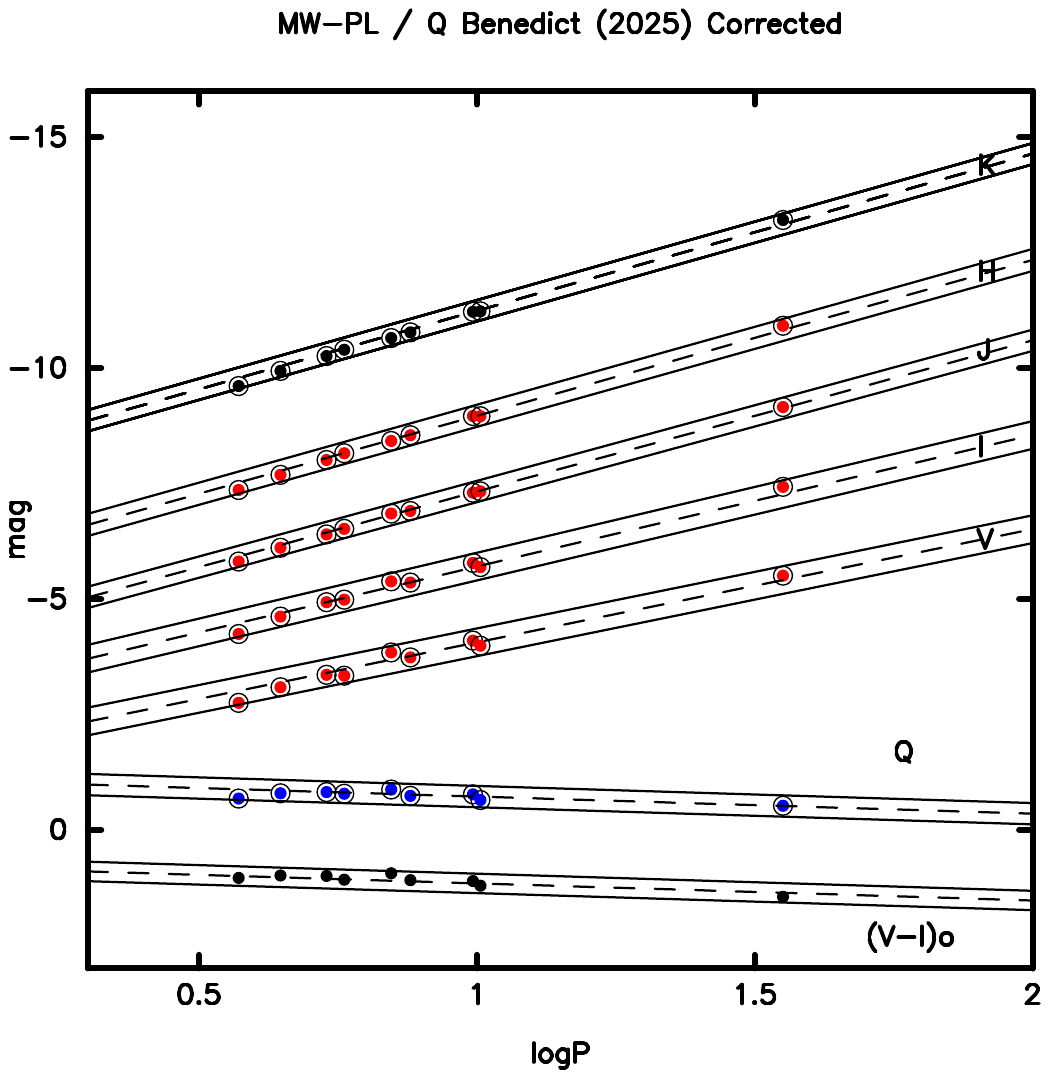} 
\caption{Same as Figure 21, except that only the HST-parallax data from Benedict et al. (2007) are shown.
}
\end{figure*}

\begin{figure*} 
\includegraphics[width=18.0cm, angle=-0]{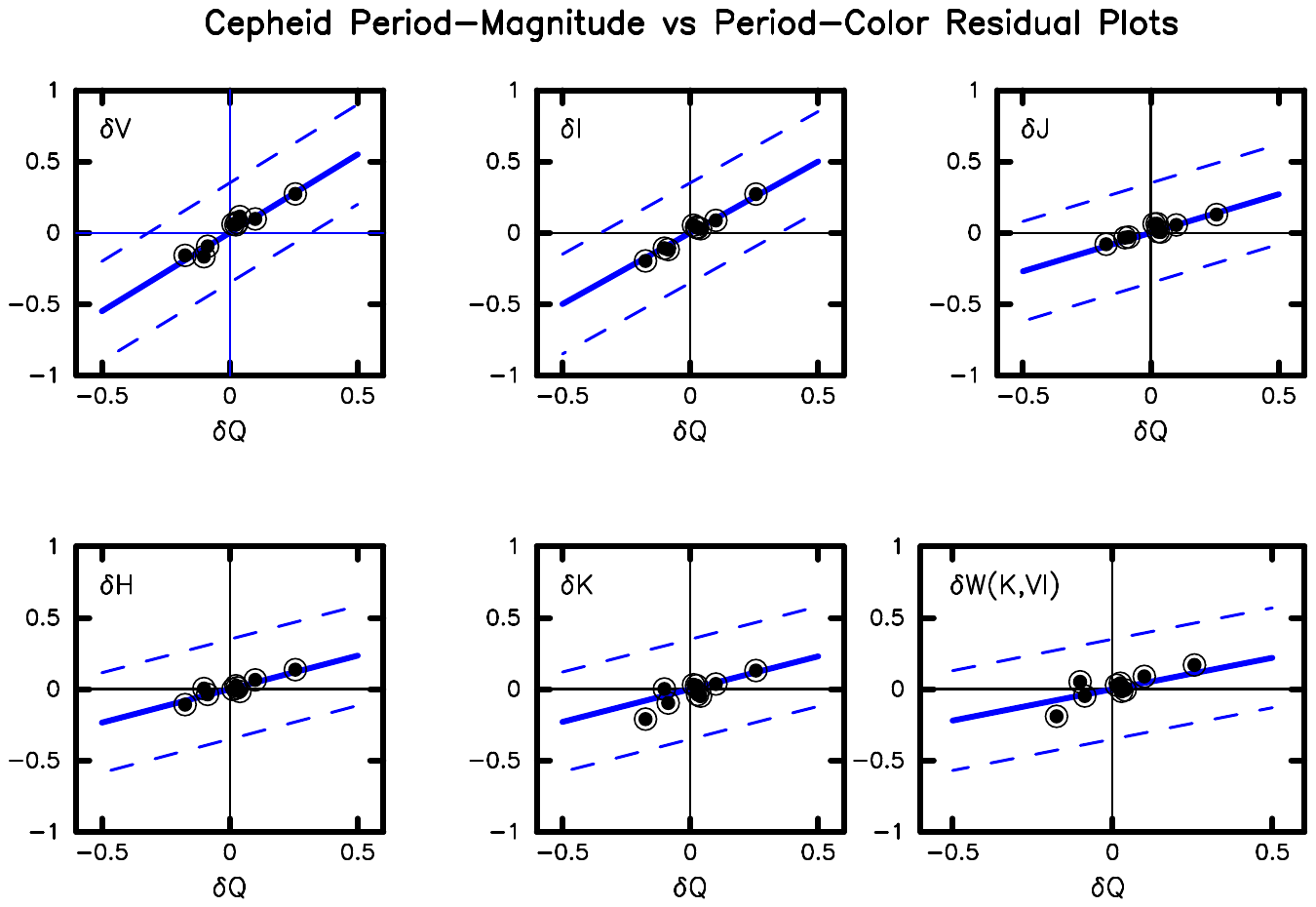} 
\caption{Same as Figure 22, except that only the HST-parallax data from Benedict et al. (2007) are shown.
}
\end{figure*}
\vfill
\subsection{Comparative Conclusions from Section 6}

The above, Section 6, has presented an independent test of Gaia parallaxes as have been published specifically for Cepheids. That comparison between HST parallaxes and Gaia parallaxes fails to reach a consensus. This offset could be signaling that one or both of the parallax zero points are in error; or that the non-overlapping samples of Cepheids selected by each of HST and Gaia have other differing systematics, (magnitude range, color range, sky distribution, to name the obvious ones) that may be better understood and possibly resolved in forthcoming releases from the Gaia consortium. 
A detailed analysis, beyond what has already been given in Madore \& Freedman (2025) is not within the scope of this paper, but this issue still needs to be ultimately considered and resolved. 
\clearpage

\section{Step 5: Baade-Wesselink Distances to Milky Way Cepheids (Fouqu\'e et al. 2007)}

We bring our PLC analysis of Cepheid distances to a penultimate close by considering an alternative method with a long and distinguished history in calibrating the Cepheid distance scale, and one that is entirely independent of geometric parallaxes. This approach exploits the physics of pulsation, using time-dependent measurements of the stellar surface motions and effective temperatures throughout the pulsation cycle to infer stellar radii and total luminosities via the Stefan–Boltzmann relation. This technique is known as the Baade–Wesselink (BW) method.

Fouqu\'e et al. (2007) present the most recent and comprehensive use of the BW Method to calibrate the PL relations for an ensemble of 59 Milky Way Cepheids observed at 7 wavelengths from the optical (BVRI) to the near infrared (JHK).
The same data analyzed in this section were the subject of an earlier investigation undertaken by Madore, Freedman \& Moak (2017) who used a variant of the technique discussed here to extract both distances and reddenings from these multi-wavelength PL relations.
Given that the results from the Baade-Wesselink analysis provide distances that are completely independent of the Gaia and/or HST direct parallaxes, we wished to re-analyze these data in exactly the same way as we have done above for the other samples.

Figure 29 shows the published absolute magnitude PL relations for 59 Milky Way field Cepheids with BW distances and individually published total line-of-sight reddenings. Here again, we note the slower than  expected decrease of the width of the PL relations with increasing wavelength, and the rather patchy cross-correlation of colors in individual Cepheids with respect to the ridge line of the two relations at the bottom of the figure, with respect to magnitude residuals from any of the PL relations. The Stefan-Boltzmann term in the theoretical PLC demands a one-to-one causal connection between these residuals, which is not obvious to the eye. Nor is the expected correlation found in any of the seven residual-residual sub-panels in Figure 30.

However, the same general pattern of coordinated deflections of individual Cepheids above and below their expected line of correlation (shown as solid blue lines) are seen in these data, just has been seen before in the other examples discussed in this paper. All of the methods are susceptible to differential distance errors. Our method for finding those errors applies equally well to BW distances as it does to Gaia or HST distances.

We have taken the average of the deflection in magnitude from the fiducial lines the seven separate bands and, as before, subtracted that average from the published absolute magnitudes plotted in Figure 29. After subtraction we have replotted the data in Figure 31. Once again this simple addition of one number systematically sharpens each of the PL relations, making their scatter significantly smaller at the longest wavelength and then increasing in the expected way and degree from near-IR to the optical. In doing so the internal ordering of the points in their respective PL relations now correlate impressively well with the star-by-star ordering in the intrinsic color plots which are untouched by the distance modulus offsets that were applied. Those (expected) correlations are shown in the seven sub-panels of Figure 32. There is still scatter in the residual-residual plots but, again it is consistent with the probable photometric errors in the input photometry, now that the parallax errors are effectively neutralized. All sources of uncertainty associated with radial-velocity measurements and the intermediate steps of the Baade–Wesselink methodology that lead to a distance determination are removed in our approach. The resulting PL relations provide both qualitative and quantitative confirmation of this improvement: they follow the theoretical expectations and are in good agreement with the corrected PL relations derived from Milky Way parallax samples, as well as with those for LMC and SMC Cepheids after accounting for depth effects.

The scatter in the correlations of magnitude and color seen in the sub-panels of Figure 32
are, respectively,  $\sigma_B= \pm 0.138,  \sigma_V = \pm 0.088, \sigma_I = \pm 0.059, \sigma_J = \pm 0.052, \sigma_H = \pm 0.073, \sigma_K = \pm 0.058$ mag. 
Using the BW parallax offsets determined here this method is now delivering distances to individual Cepheids in this Milky Way sample at the $\pm$ 3-7\% (filter dependent) level per star.

Five different samples of Cepheids and three different methods of distance determinations all converge on highly precise period-luminosity-color relations aligning with theoretical expectations.

\begin{figure*} 
\includegraphics[width=18.0cm, angle=-0]{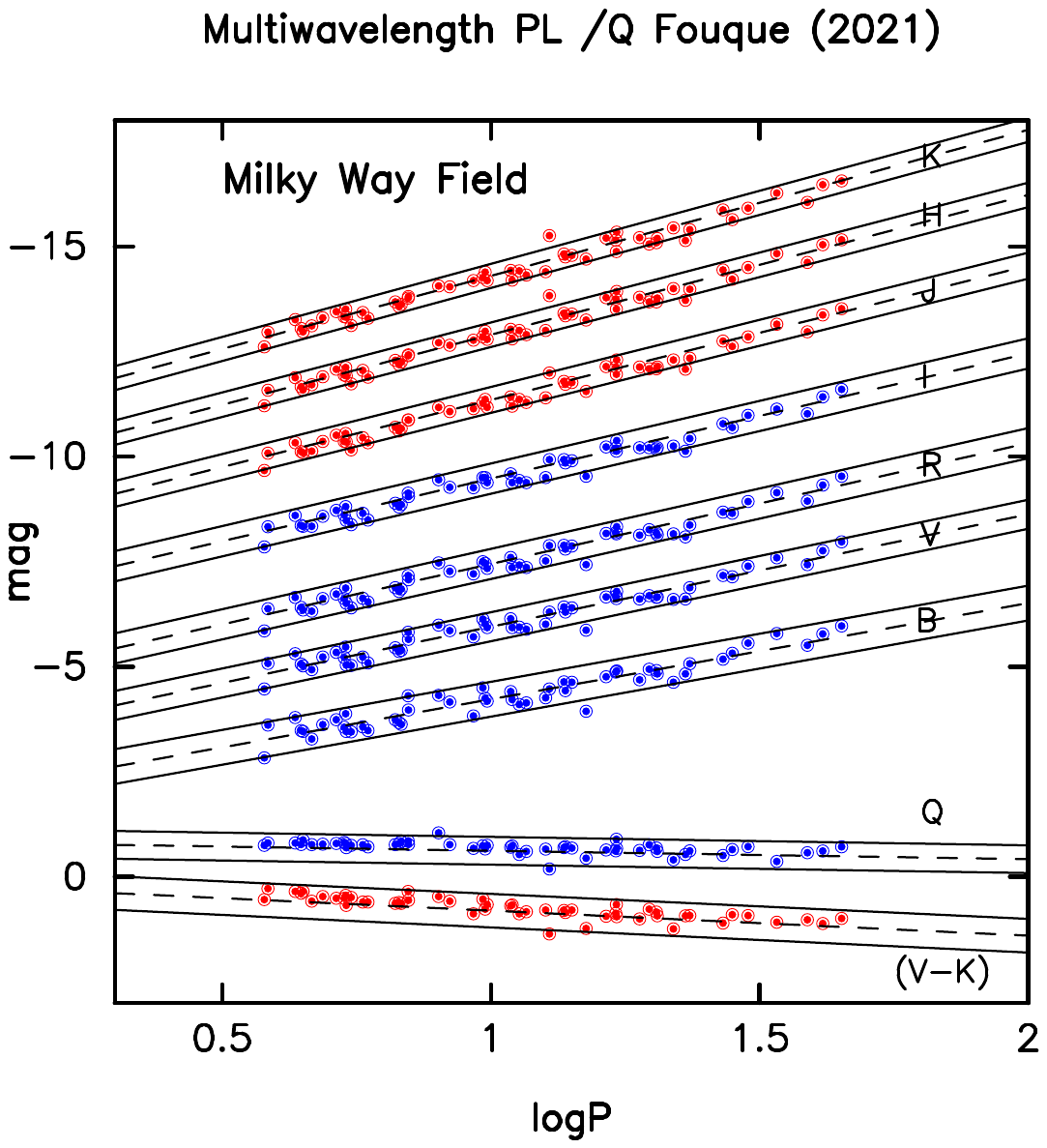} 
\caption{Intrinsic period-luminosity (BVRI,JHK: blue and red points, resp.) and period-color relations (Q, blue points and (V-K), red points, below) as taken from Fouqu\'e et al. (2007). Individual relations are offset for clarity by forcing reduced overlap.
}
\end{figure*}

\begin{figure*} 
\includegraphics[width=18.0cm, angle=-0]{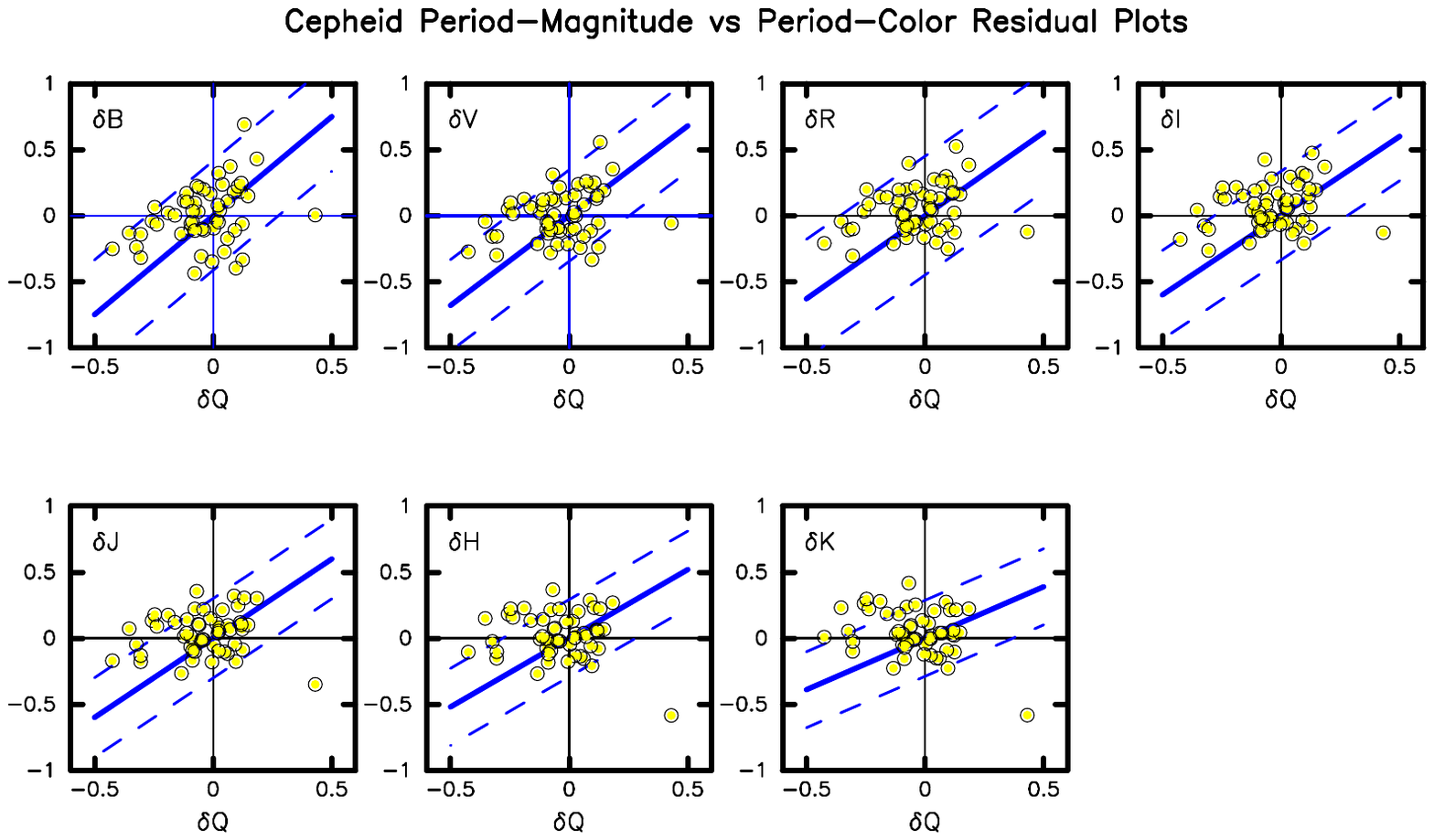} 
\caption{Magnitude and color deviations from the Period-Luminosity and the Reddening-Free (Q) Period-Color relations  as a function of filter wavelength (BVRI,JHK) shown in separate panels from left to right and top to bottom. Details are the same as in Figure 1. 
}
\end{figure*}
\clearpage
\begin{figure*} 
\includegraphics[width=18.0cm, angle=-0]{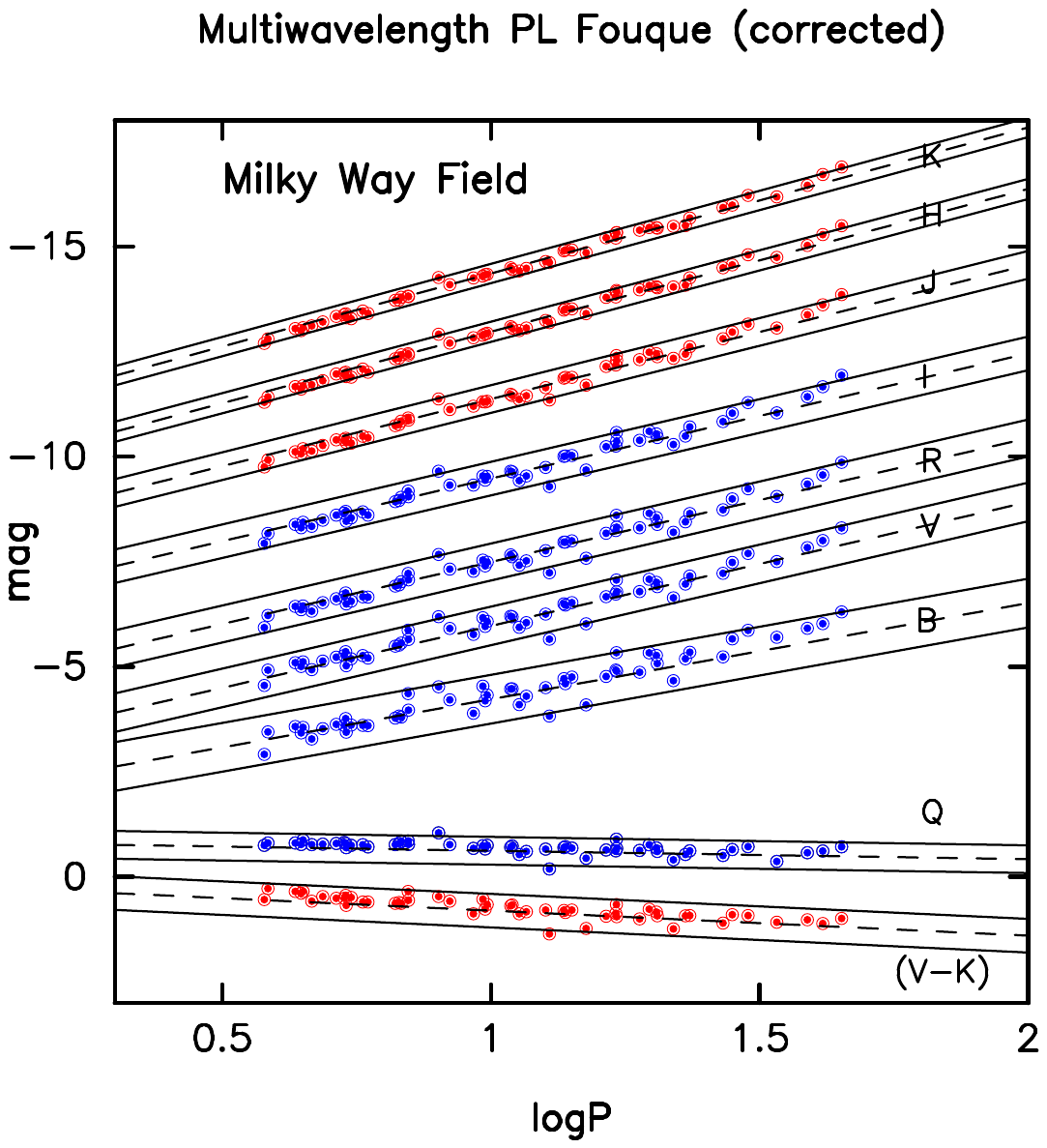} 
\caption{Intrinsic Period-Luminosity (BVRI,JHK: blue and red points,resp.) and Period-Color relations (Q, blue points and (V-K), red points, below) as taken from Fouqu\'e et al. (2007). Individual relations are offset for clarity by forcing reduced overlap.
}
\end{figure*}

\clearpage
{\subsection{}

\begin{figure*} 
\includegraphics[width=18.0cm, angle=-0]{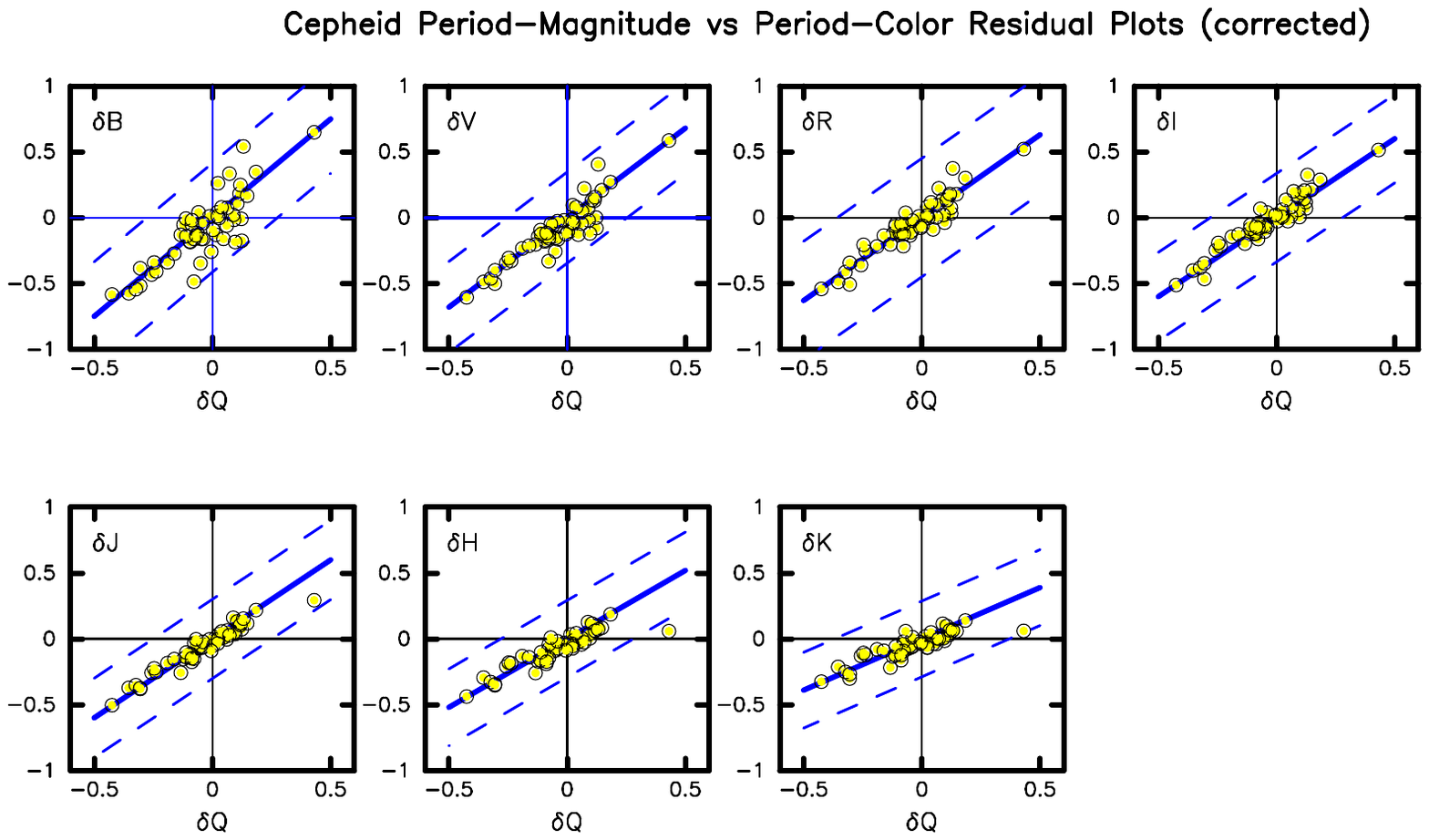} 
\caption{After correlated distance modulus offset corrections, magnitude and color deviations from the Period-Luminosity and the reddening-free (Q) Period-Color relations are shown as a function of filter wavelength (BVRI,JHK) in separate panels from left to right and top to bottom. Details are the same as in Figure 1. 
}
\end{figure*}

\section{Step 6: The Complete Gaia Milky Way Cepheid Sample}

We now turn to the Ripepi et al. (2019) sample of Milky Way Cepheids based on the DR3 release. Although the Gaia observations discussed here are restricted to only two broad-band filters (GB and GR), they do include virtually all known Galactic Cepheids having individually and independently determined reddenings. Furthermore, all of the photometry is on one system, and the mean magnitudes and colors for the Cepheids are based on densely-sampled, high signal-to-noise light curves and updated periods. 

All that said, it is obvious that a comparison of the scatter in the two intrinsic  (GB and GR) PL relations for the MW sample (with an average distance of about 6 kpc) are quite disappointing in comparison to the extremely well-defined and much narrower PL relations found in DR3 for Cepheids in the LMC and SMC, at distances of 50 and 63 kpc, 10 times further than the MW sample, and 100 times fainter.

The published parallax-corrected and extinction-corrected Gaia Milky Way PL relations are shown in Figure 33.  The published parallax errors, expressed as equivalent (but naturally asymmetric) distance-modulus errors are shown as thin black lines.  In Figure 34 we show the two residual-luminosity versus residual-color plots for the GB and GR data. The scatter is so large as to hide any indication of a correlation between the two quantities. But there is, once again, a surprisingly high degree of order in the apparent chaos of these ``scatter plots". Three arrows in the plot illustrate the point that each magnitude deviations away from the fiducial blue lines is identical across both wavelengths. We conclude that the scatter in the PL relations is once again due to (achromatic) parallax errors in the Milky Way Cepheid sample.

Averaging the two magnitude residuals for each of the Cepheids, and subtracting that average from their published absolute magnitudes, we replot the parallax-corrected PL relations in Figure 35. The result is impressive. Not only is the scatter now just as small as the LMC scatter, but it increases, as hoped, from the long wavelength GR band to the shorter GB band PL relation. Finally, it has to be remarked that the cross-correlation of the ordering of Cepheids in the intrinsic $(GB-GR)_o$  period-color relation is reflected in the magnitude ordering of the Cepheids across the instability strip as seen in both of the parallax-corrected PL relations. That is highly unlikely to be an accident.
And it is not due to fine tuning. It is the result of a single correction being applied to each Cepheid regardless of the bandpass
it is being observed in. At the top of the plot we show the PLC fit to the data, where the resulting scatter is only $\pm0.072$ mag. This result and its implications for the Cepheid distance scale will be discussed in more detail in Paper II. 


\subsection{Are the Parallax Corrections Derived Here Reasonable?}

Are the  errors attributed to parallaxes in the preceding sections plausible? In addition to the  
theoretically predicted results in the refining of the monotonic behavior of the scatter in the multi-wavelength PL relations as a function of wavelength, and in the coincidentally fine structure correlations of the placement of individual stars within both the PL and PC relations, we have the following independent test of credibility. In Figure 37 we plot our distance modulus corrections as a function of GB magnitude (yellow points), shown in the foreground overlaying the published symmetric 2-sigma parallax errors (black points) expressed as asymmetric distance modulus uncertainties, in the background (see caption for details). The growing dispersion as a function of increasing magnitude flows is as expected, both qualitatively and quantitatively. We take this plot as evidence that the corrections we are applying to the Gaia parallaxes are real, and that they are quantitatively correct in both their sign and magnitude.

\begin{figure*} 
\includegraphics[width=18.0cm, angle=-0]{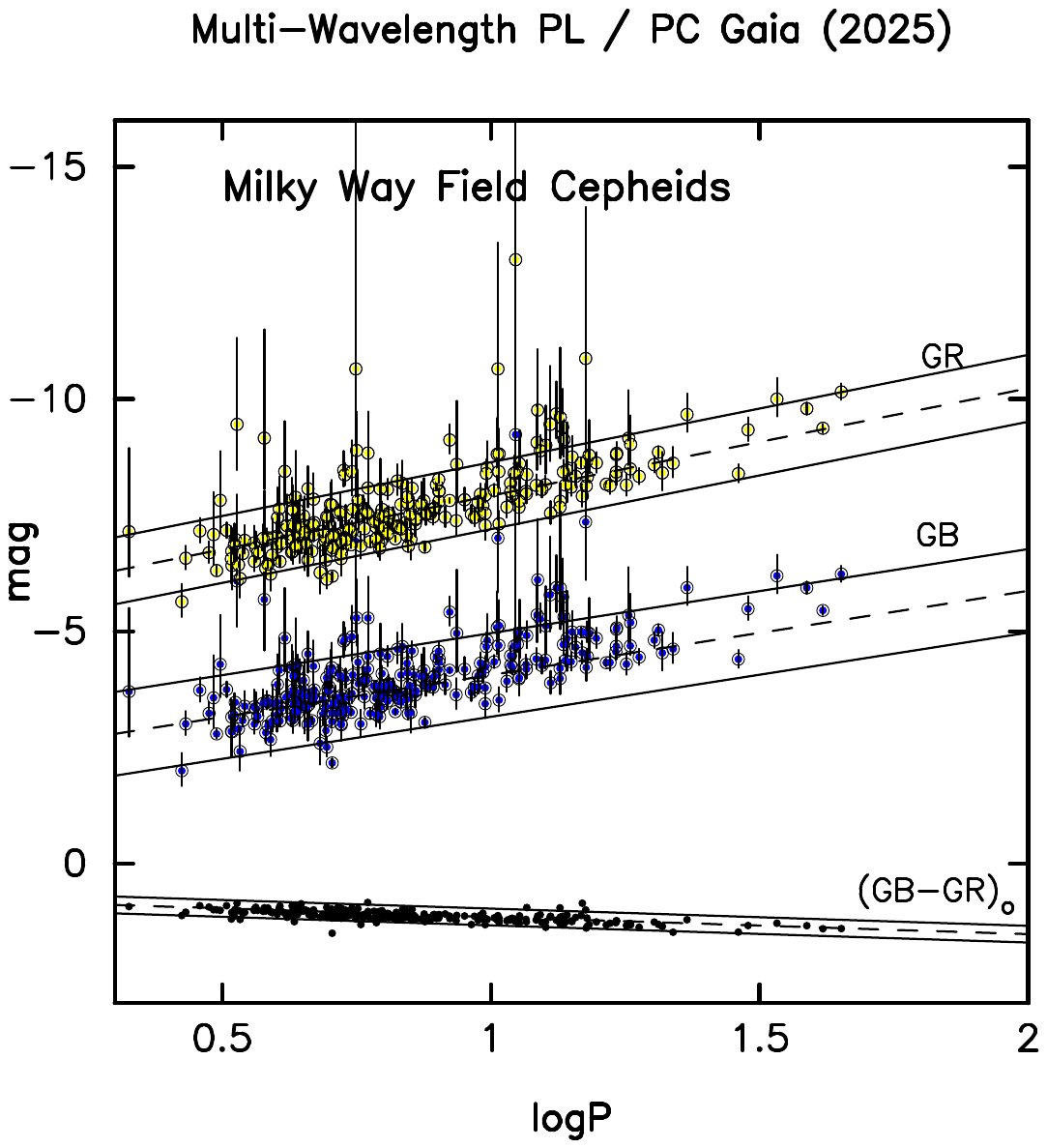} 
\caption{The two Gaia GB (blue) and GR (yellow) Period-Luminosity relations (with one-sigma, published parallax errors) are shown in the middle of the panel, and the star-by-star, reddening-corrected Period-Color relations $(GB-GR)_o$are shown as black points at the bottom.
}
\end{figure*}

\begin{figure*} 
\includegraphics[width=18.0cm, angle=-0]{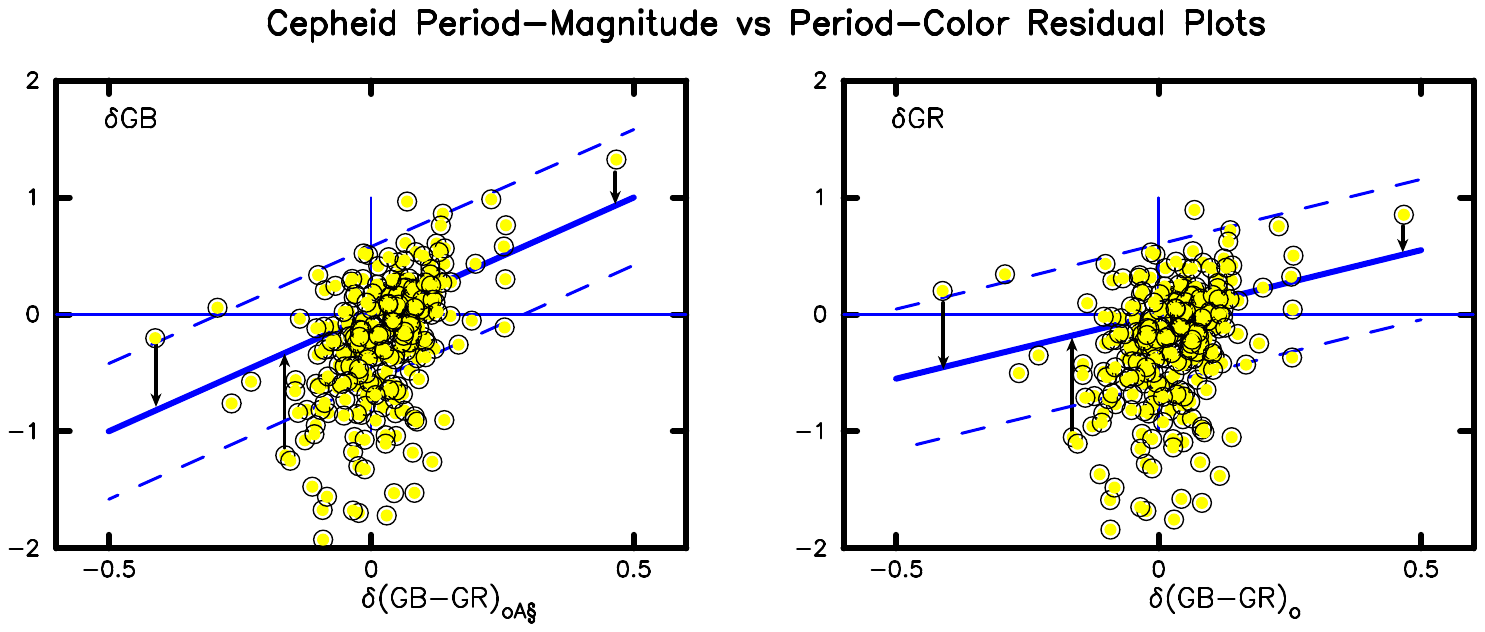} 
\caption{Magnitude and color deviations from the Period-Luminosity and the reddening-corrected $(GB-GR)_o$ Period-Color relations, respectively.
GB and GR plots are shown in the left and right sub-panels. Black, vertical arrows show three examples of magnitude residuals that are highly correlated in sign and amplitude. They are achromatic and interpreted to be (now measurable) parallax errors.
}
\end{figure*}
\begin{figure*} 
\includegraphics[width=18.0cm, angle=-0]{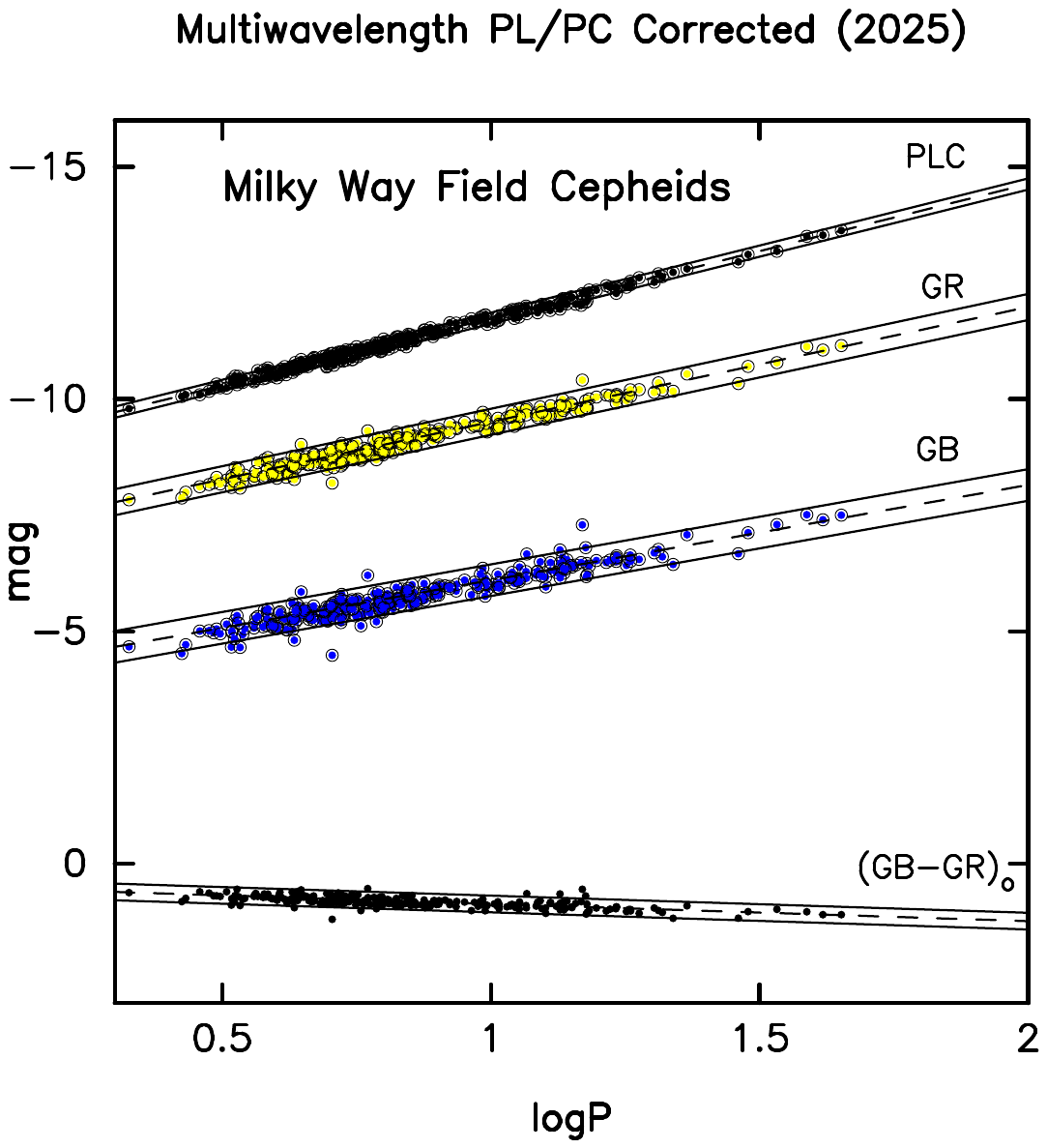} 
\caption{After correlated distance modulus offset corrections have been applied, the two Gaia GB (blue) and GR (yellow) Period-Luminosity relations are shown in the middle of the panel, and the (unaffected) star-by-star, reddening-corrected Period-Color relations $(GB-GR)_o$are shown as black points at the bottom. At the top of the frame is the Period-Luminosity-Color fit to the corrected data (black points).
}
\end{figure*}

\begin{figure*} 
\includegraphics[width=18.0cm, angle=-0]{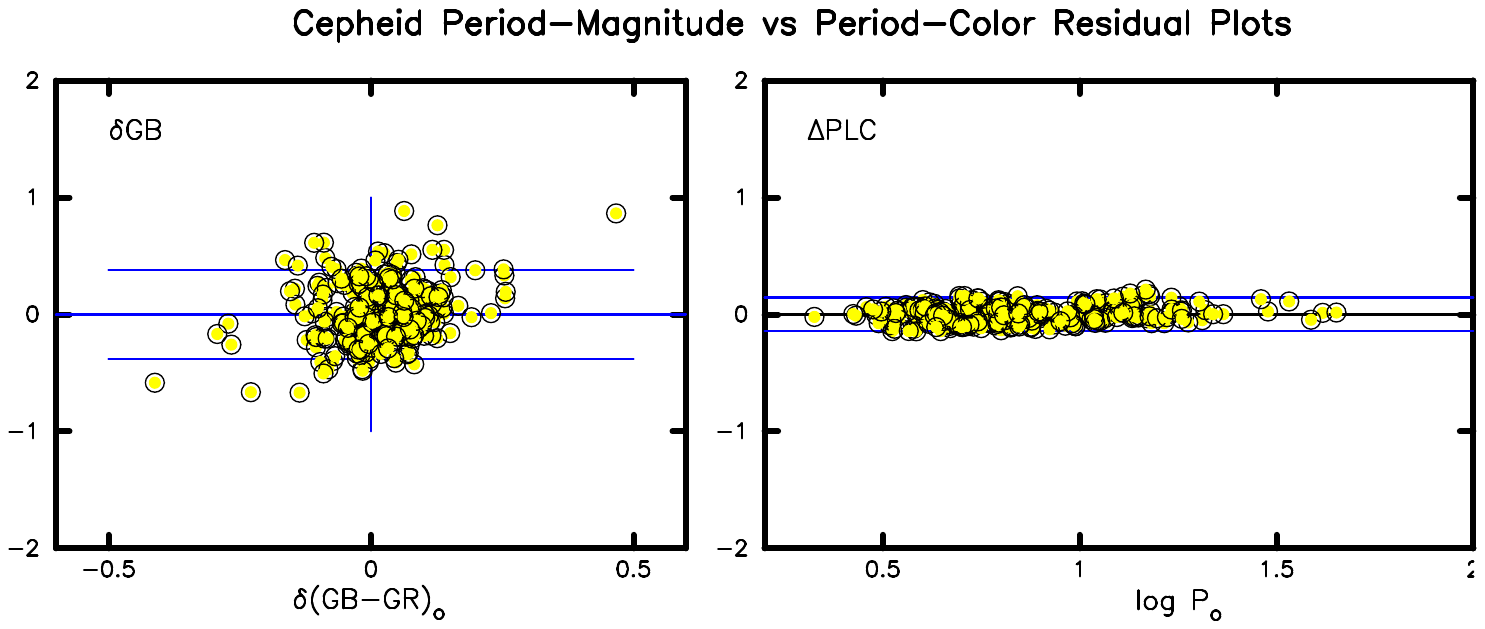} 
\caption{After correlated distance-modulus offset corrections have been applied  the residual magnitude deviations from the PLC relation are shown plotted as a function of the intrinsic color $(GB-GR)_o$ in the left panel, and as a function of the period in the right panel. The scatter is $\pm$0.074~mag. For comparison Ripepi et al. (2022) report one-sigma scatters for parallax-uncorrected $GB$ and $GR$ PL relations of $\pm$0.51 and 0.31~ mag, respectively. These are {\bf factors of 7 and 4 times larger} than the corrected values used in the PLC solution found in the present study.
}
\end{figure*}

\begin{figure*} 
\includegraphics[width=18.0cm, angle=-0]{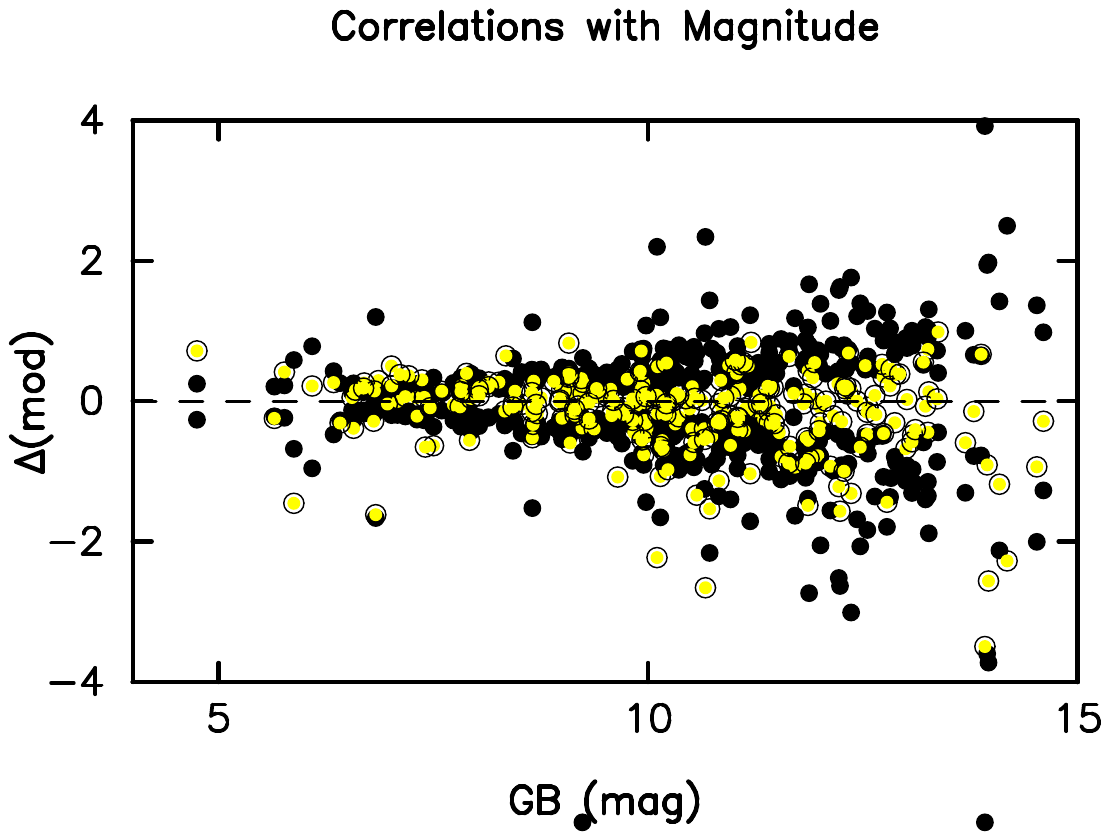} 
\caption{Distance modulus correction, $\Delta$(mod), as determined here, plotted as a function of Gaia magnitude ($GB$ in yellow) showing not only the expected increase as a function of the Cepheid's magnitude, but any other position-dependent terms which may account the relatively fixed scatter at the bright, magnitude 5 to 7 range. Black circles in the background are the $\pm$ two-sigma errors in the parallaxes, transformed to  asymmetric errors in the distance moduli, showing overall consistency of the measured distance moduli corrections and the calculated/predicted uncertainties, as published.}
\end{figure*}

\section{Just How Precise Are the New Parallaxes?}

Figure 37 shows the photometric precision of our newly corrected distance moduli in the form of vertical scatter measured in magnitudes. Taking the absolute value of that deviation on a star-by-star basis and converting it into a parallax error at the new distance of the Cepheid we can compare that error with the error published by the Gaia project. The average gain in precision is a factor of 19.3x; however, this number is dominated by outlying values in a very extended tail. Cutting the sample to only those with a ratio below 100 drops the mean ratio 6.1x. But, perhaps a better estimator is the median of the distribution, which gives a gain of at least a factor of 1.95x for half of the Cepheids studied here. But even that reduced value is still considerable, given that at least 133 Cepheids have corrections in parallax precision in excess of 195\%. Nevertheless, Figure 37 shows that virtually all of these individually applied corrections are within the bounds of the parallax uncertainties as published in DR3.

\clearpage
\section{Concluding Remarks}


Informed by first-principles physics, we have made the case that the luminosities of Cepheid variables must be minimally dependent upon at least two independent variables. From the theoretical/astrophysical perspective those two variables are (a) the physical radius of the stellar atmosphere and (b) the surface temperature at that same radius. Bolometric luminosities then become uniquely defined by geometry and thermodynamics. This application of the Stefan-Boltzmann equation to spherical, self-luminous objects applies to all stars, including Cepheids, in the mean and/or at any phase in their pulsation cycles. Mapped into observationally accessible quantities, while maintaining the same minimal dimensionality, those two control parameters find proxies in the pulsation period and the intrinsic color of the star. That specific combination of observables leads to a period-luminosity-color (PLC) relation. The one-parameter (marginalized) PL relation fails to capture the full physical complexity of Cepheid variability and has, for decades, promoted oversimplifications that reduce the attainable precision of the extragalactic distance scale.

The intrinsic PLC relation can, of course, only be used if the apparent magnitudes and colors of individual stars/Cepheids are each corrected for total line-of-sight reddening. In our effort to calibrate the PLC relation locally using Milky Way Cepheids with published individual reddenings and geometric parallaxes from HST and Gaia, the resulting color–luminosity relation exhibited a level of scatter far exceeding that expected from the intrinsic width of the instability strip, indicating that the dominant contribution is observational rather than physical in origin.

We have determined the source of that scatter by observing that the deviations perpendicular to the color plane (i.e. the magnitude deviations for any given Cepheid) are of the same degree and of the same sign, independent of the wavelength of the PL relation being considered. That is, the primary source of the residual scatter is achromatic. This leads to the obvious conclusion that errors in the parallaxes are the underlying cause. 

Having detected these individual deflections, it allows us to measure them, and thereby correct the parallaxes on an individual basis. These signed corrections to the individual stars correlate in magnitude with the independently published Gaia standard errors to the same individual Cepheid parallaxes, giving credence to our causal explanation.

Do these corrections make a difference? Yes, in two very important ways. Applying {\it a single correction} to a given Cepheid's absolute magnitude for one of its observed multi-wavelength PL relations, and then re-deriving and re-plotting the PL relations, gives rise to the fully expected decrease of scatter as a function of increasing wavelength. Moreover, the remaining scatter in these parallax-error corrected PL relations now correlate (again, at all wavelengths) with the color disposition of that Cepheid in the instability strip. This brings the $rms$ scatter on the final PLC fit to the data down to the level of the photometric precision of the input colors and apparent magnitudes: the final scatter is found typically to be $\pm$0.04~mag, giving a distance precision $\pm$2\% per Cepheid, galactic or extragalactic.

A more detailed look at the implications of an extinction and  reddening corrected PLC (derived for a wider range of filters, including the MIR) applied to the extragalactic distances scale, will be presented in a forthcoming companion paper (Freedman et al. 2026).

\clearpage
~~~~~~~~~~~~~~~~~~~~~~~~~~~~~~~~~~~~~~~~~~~~~~~~~~~~~~~~~~ACKNOWLEDGMENTS
\medskip

 We thank the University of Chicago 
and Observatories of the Carnegie Institution for Science for their support of our long-term research incrementally allowing a better understanding of the expansion rate of the Universe and the systematics currently confounding those determinations. Support for this work was also provided in part by NASA through grant number HST-GO-13691.003-A from the Space Telescope Science Institute, which is operated by AURA, Inc., under NASA contract NAS~5-26555. No AI software was used at any point in the execution of this research. Finally, our thanks to Dan Majaess for early comments on the manuscript; and special thanks to Giuseppe Bono and Vincenzo Ripepi for their individual insightful comments and additions just prior to submission. 
\section{References}
\par\noindent
Andrievsky, S.M. \&  Koutyuch, V.V. 2011, AJ, 142, 51
\par\noindent
Baade, W. 1956, PASP, 68, 5
\par\noindent
Benedict, G.F., McArthur, B.E., Feast, M.W. et al. 2007, AJ, 133, 181
\par\noindent
Bono, G., Caputo, F., Castellani, V. \& Marconi, M. 1999, ApJ, 512, 711
\par\noindent
Bono, G. \& Marconi, M. 1999, IAU Symp. 190, 527
\par\noindent
Breuval, L., Kervella, P., Anderson, R., et al. 2020, A\&A, 643, 115 
\par\noindent
Breuval, L., Kervella, P., Wielg\'orski, P., et al. 2021, ApJ, 913, 38 
\par\noindent
Caldwell, J.A.R. \& Coulson, I.M.1986, MNRAS, 218, 223
\par\noindent
de Vaucouleurs, G. 1983, MNRAS, 202, 367
\par\noindent
di Valentino, E. et al. 2021, Astropart. Phys. 131, 102605 
\par\noindent
Fouqu{\'e}, P., Arriagada, P, Storm, J., et al. 2007, A\&A, 476, 73  
\par\noindent
Graczyk, D., Pietrzy\'nski, G., Thompson, I.B. et al. 2020, ApJ, 904, 13
\par\noindent
Haschke, R., Grebel, E. K., \& Duffau, S. 2012, A\&A, 543, A106
\par\noindent
Hiltner, W.A. \& Johnson, H.L. 1956, ApJ, 124, 367
\par\noindent
Lamers, H.J.G.L.M. \& Levesque, E.M. 2017, Understanding Stellar Evolution, IOP Publishing, Bristol, UK 
\par\noindent
Lesmasle, B, Hajdo, G., Koutyukh, L., et al. 2018, A\&A, 618, 610
\par\noindent
Lindegren, L., Klioner, S.A., Herna\'ndez, J., et al. 2021 A\&A, 649, A2 
\par\noindent
Lucchini, S. 2024, Ap\&SS, 369, 114
\par\noindent
Luck, R.E. \& Lambert, D.L. 2011, AJ, 142, 136
\par\noindent
Luck, R.E. 2018, AJ, 156, 171
\par\noindent
Madore, B.F., Freedman, W.F. \& Moak, S. 2025, ApJ, 842, 42 
\par\noindent
Madore, B.F. \& Freedman, W.F. 2025, ApJ, 983, 161
\par\noindent
Madore, B.F., Freedman, W.F. \& Owens, K. 2025 ApJ, 981, 32
\par\noindent
Menendez-Delgado, J.E., Amayo, A., Arellano-Cordova, K.Z. et al. 2022, MNRAS, 510, 4436
\par\noindent
Persson, S.E., Madore, B.F., Kzremi\'nski, W., et al. 2004, AJ, 128, 2239
\par\noindent
Ripepi, V., Molinaro, R., Musella, I. et al. 2019, A\&A, 615, 14
\par\noindent
Sandage. A.R., 1972, QJRAS, 13, 202
\par\noindent
Sandage. A.R. \& Tammann, G.A., 1968, ApJ, 151, 532
\par\noindent
Sandage, A. \& Tammann, G. A. 1982, ApJ, 256, 339
\par\noindent
Smith, R. W. 1982, Cambridge University Press 
\par\noindent
Smitha, S. \& Annapurni, S., 2012, ApJ, 744, 128
\par\noindent
van der Marel, R.P. \& Cioni, M.-R.L. 2001, AJ, 122, 1807

\clearpage
\end{document}